\documentclass[rsi,amsmath,amssymb,graphicx,twocolumn]{revtex4-1}

\usepackage{graphicx}
\usepackage{graphics}
\usepackage[font=small]{subcaption}
\usepackage{xspace}
\usepackage[export]{adjustbox}
\usepackage{hyperref}
\hypersetup{
    colorlinks=true,
    linkcolor=blue,
    filecolor=blue,      
    urlcolor=blue,
    citecolor=blue
}
\usepackage[capitalize]{cleveref}
\usepackage{xcolor}

\newcommand{\SI}[2]{#1~\mathrm{#2}}
\newcommand{\Ip}{I_\mathrm{P}}
\newcommand{\Bt}{B_\mathrm{T}}
\newcommand{\Bp}{B_\mathrm{P}}
\newcommand{\psin}{\psi_\mathrm{N}}
\newcommand{\sqrtpsin}{\sqrt{\psin}}
\newcommand{\thetap}{\theta_\mathrm{p}}
\newcommand{\Ec}{E_\mathrm{c}}
\newcommand{\vpar}{v_\parallel}
\newcommand{\vperp}{v_\perp}

\newcommand{\mywidth}{0.2\textwidth}
\newcommand{\myheight}{7cm}

\newcommand{\add}[1]{\textcolor{red}{#1}}
\renewcommand{\add}[1]{#1\unskip}
\newcommand{\sub}[1]{\sout{#1}}
\renewcommand{\sub}[1]{\unskip}

\newcommand{\reved}[1]{\textcolor{red}{#1}}
\renewcommand{\reved}[1]{#1\unskip}
\newcommand{\revone}[1]{\textcolor{magenta}{#1}}
\renewcommand{\revone}[1]{#1\unskip}
\newcommand{\revtwo}[1]{\textcolor{cyan}{#1}}
\renewcommand{\revtwo}[1]{#1\unskip}

\newcommand{\rsiadd}[1]{\textcolor{red}{#1}}
\renewcommand{\rsiadd}[1]{#1\unskip}

\begin{document}

%\preprint{}

\title{Synthetic measurements of runaway electron synchrotron emission in the SPARC tokamak} %Title of paper



%\author{}
%\thanks{}
%\affiliation{}

\author{R.A.~Tinguely}
\email[]{tinguely@psfc.mit.edu}
%\affiliation{MIT Plasma Science and Fusion Center, Cambridge, MA, USA}
\affiliation{\rsiadd{Plasma Science and Fusion Center, Massachusetts Institute of Technology, Cambridge, MA 02139, USA}}

%\author{A.M.~Rosenthal, M.~Silva Sa, M.~Jean, I.~Abramovic}
%\affiliation{\rsiadd{Commonwealth Fusion Systems, Devens, MA  01434, USA}}
%\affiliation{Commonwealth Fusion Systems, Devens, MA, USA}

\author{A.M.~Rosenthal}
\affiliation{Commonwealth Fusion Systems, Devens, MA  01434, USA}

\author{M.~Silva Sa}
\affiliation{Commonwealth Fusion Systems, Devens, MA  01434, USA}

\author{M.~Jean}
\affiliation{Commonwealth Fusion Systems, Devens, MA  01434, USA}

\author{I.~Abramovic}
\affiliation{Commonwealth Fusion Systems, Devens, MA  01434, USA}

%\collaboration{SPARC Team}

\date{\today}

\begin{abstract}
    With plasma currents up to 8.7~MA, the SPARC tokamak runs the risk of forming multi-MA beams of relativistic ``runaway'' electrons (REs) which could damage plasma facing components if unmitigated. The infrared (IR) and visible imaging and \add{visible} spectroscopy systems in SPARC are designed with measurements of synchrotron emission from REs in mind. Synchrotron radiation is emitted by REs along their direction of motion, opposite the plasma current. Matched clockwise and counterclockwise wide views \add{are proposed to} detect synchrotron and background radiation, allowing observation of RE synchrotron emission in both plasma current configurations. Due to SPARC’s high toroidal magnetic field strength, 12.2~T on axis, the synchrotron light spectrum is expected to peak in the visible-IR wavelength range. The synthetic diagnostic tool SOFT \reved{(Synchrotron Orbit-Following Toolkit)} is used to model synchrotron images and spectra for three scenarios, \add{with appropriate magnetic equilibria for each}: REs generated during plasma current ramp-up, steady-state flat-top (although unlikely, but serving as a reference), and disruptions. \revtwo{Required time resolutions, achievable spatial coverage, and appropriate spectral ranges for various RE energies are assessed.} %Temporal, spatial, and energy/spectral resolutions are assessed.
\end{abstract}

%With plasma currents up to 8.7~MA, the SPARC tokamak runs the risk of forming multi-MA beams of relativistic ``runaway'' electrons (REs) which could damage plasma facing components if unmitigated. The infrared/visible imaging and spectroscopy systems in SPARC are designed with measurements of synchrotron emission from REs in mind. Synchrotron radiation is emitted by REs along their direction of motion, opposite the plasma current. Matched clockwise and counterclockwise wide views will detect synchrotron and background radiation, allowing observation of RE synchrotron emission in both plasma current configurations. Due to SPARC’s high toroidal magnetic field strength, 12.2~T on axis, the synchrotron light spectrum is expected to peak in the visible wavelength range, even for low-energy REs. The synthetic diagnostic tool SOFT is used to model synchrotron images and spectra for three scenarios: REs generated during plasma current ramp-up, steady-state flat-top (although unlikely, but serving as a reference), and disruptions; appropriate magnetic equilibria are used in each case. Scans are performed in RE energy, pitch angle, and minor radial location. Temporal, spatial, and energy/spectral resolutions are assessed.

%\pacs{}% insert suggested PACS numbers in braces on next line

\maketitle %\maketitle must follow title, authors, abstract and \pacs

%\section{}
%\label{}
%\subsection{}
%\subsubsection{}

%\input{outline}
\section{Introduction}\label{sec:intro}

    \revtwo{
        The SPARC tokamak is a compact fusion device whose construction is progressing \add{rapidly in} Devens, MA. SPARC has major and minor radii $R_0 = \SI{1.85}{m}$ and $a = \SI{0.57}{m}$, respectively \cite{Creely2020}. Its high on-axis magnetic field strength, $B_0 = \SI{12.2}{T}$, is enabled by high temperature superconducting technology \cite{Hartwig2024}. The main mission of SPARC is to demonstrate fusion breakeven, $Q > 1$, in a magnetic confinement device; the highest power ``Primary Reference Discharge'' (PRD) is expected to achieve $Q \sim 10$.
    }
    %Construction of the SPARC tokamak \cite{Creely2020} and site is progressing \add{rapidly in} Devens, MA. In parallel, the development of diagnostics for early operations is maturing quickly \cite{Reinke2024}. Among them are the imaging and spectroscopy systems \cite{SilvaSa2024,Rosenthal2024}, with requirements including  machine inspection and protection. This paper will focus on the detection of high-energy runaway electrons (REs) via their synchrotron emission in the infrared (IR) and visible light spectrum. 
    
    In parallel with the tokamak and site, the development of diagnostics for early operations is maturing quickly \cite{Reinke2024}. Among them are the imaging and spectroscopy systems \cite{SilvaSa2024,Rosenthal2024}, with requirements including  machine inspection and protection. This paper will focus on the detection of high-energy runaway electrons (REs) via their synchrotron emission in the infrared (IR) and visible light spectrum. 

    In a plasma, electrons can be continuously accelerated to relativistic speeds when an electric field $E$ overcomes collisional friction (among other energy loss mechanisms). Plasma conditions must satisfy \add{$E/\Ec \gg 1$}, with the critical field $\Ec\mathrm{[V/m]} \approx 0.005 \ln\Lambda \, n_{e}\mathrm{[10^{20} m^{-3}]}$ \cite{Connor1975}, where $\ln\Lambda$ is the Coulomb logarithm \add{and $n_e$ is the electron density}. This situation can arise during (i)~plasma start-up, as both the plasma current $\Ip$ and density $n_e$ ramp up ($E/\Ec \sim 100$), and (ii)~plasma disruptions, when the fast current quench induces a large loop voltage ($E/\Ec \sim 1000$). REs can also form during the steady-state $\Ip$ ``flat-top'' at low $n_e$; however, this scenario is often avoidable with proper density control.

    Interestingly, electron cyclotron emission becomes relativistic ``synchrotron'' emission \add{for REs} when \add{their} energies exceed \add{their} rest mass, $\SI{0.511}{MeV}$. As a sum of many cyclotron harmonics, this effectively continuous synchrotron spectrum is shifted from microwave to IR and visible wavelengths. In contrast to \add{much of} plasma radiation which is isotropic, the angular distribution of synchrotron emission is highly forward-directed, along the RE's velocity vector. This means that, barring reflections, synchrotron light is only observed \add{coming} from the counter-$\Ip$ direction. 

    \revtwo{
        The purpose of this measurement is primarily twofold: First, if REs are unintentionally generated in SPARC, this passive measurement can help inform tokamak operations to avoid high energy, high current RE beams, which could possibly damage plasma facing components. Second, dedicated physics experiments are planned to study plasma disruptions and mitigation, including Massive Gas Injection (MGI) as well as the Runaway Electron Mitigation Coil (REMC) \cite{Sweeney2020,Tinguely2021REMC}.
    }
    
    Table~1 in \cite{Tinguely2018} provides an extensive list of tokamaks with measurements of RE synchrotron emission. The rest of this paper will explore capabilities and opportunities for similar measurements in SPARC: Proposed in-vessel views and associated optics 
        %\footnote{Disclaimer: No optical specifications given in this work should be considered final.}
    are discussed in \cref{sec:diagnostic}. Then, synthetic synchrotron emission is simulated for three plasma phases: flat-top in \cref{sec:flattop}, serving as a reference case; start-up in \cref{sec:startup}; and a disruption in \cref{sec:disruption}. Finally, a summary is provided in \cref{sec:summary}. Importantly, note that optical specifications presented here are \emph{not yet finalized} in the diagnostic design and should therefore be treated as notional. 
    \reved{Many factors will affect the final design such as electromechanical forces or radiation limits which may effect locations and sizes of the aperture and mirrors.}
    %\reved{Many factors will affect the final design, such as electromechanical forces on aperture/mirror locations or radiation limits on penetration/mirror sizes.}

\section{Proposed in-vessel optics}\label{sec:diagnostic}

    While many in-vessel views are planned for imaging and spectroscopy in SPARC, this paper will focus on two matched clockwise (CW) and counterclockwise (CCW) wide-angle views, co-located at the same midplane port and shown in \cref{fig:camera-views}. These see much of the main chamber, including the outer limiters, ICRH antennas, and inner wall. Nominal directions of $\Ip$ and toroidal magnetic field $\Bt$ are CW when viewing the tokamak from above, which means REs will move in the CCW direction. Synchrotron emission would therefore be seen on the ``left'' side of the tokamak, i.e. \cref{fig:cam-view-cw}, so all following analyses are carried out for this CW field of view. Note that the CCW view can be used for (i)~background subtraction and (ii)~synchrotron measurements in reversed $\Ip/\Bt$ configurations (which preserve helicity). \revone{Limitations on the former may arise due to the different features on the outer wall within each view, e.g. perhaps causing different reflections; however, this can be assessed in actual operations, and some kind of spatial averaging of the background could also be performed.}
    
    %\revone{Note that the CCW view can be used for synchrotron measurements in reversed $\Ip/\Bt$ configurations (which preserve helicity). Background subtraction can also be attempted with the CCW view, although the outer wall does not have identical features in both CW/CCW views; limitations, e.g. from reflections, can be assessed in actual operations.}
    
    %Note that the CCW view can be used for (i)~background subtraction and (ii)~synchrotron measurements in reversed $\Ip/\Bt$ configurations (which preserve helicity). 
 
    \renewcommand{\mywidth}{0.49\columnwidth}
    \begin{figure}%[h!]
        \centering
        \begin{subfigure}{\mywidth}
            \centering
            \includegraphics[height=4cm]{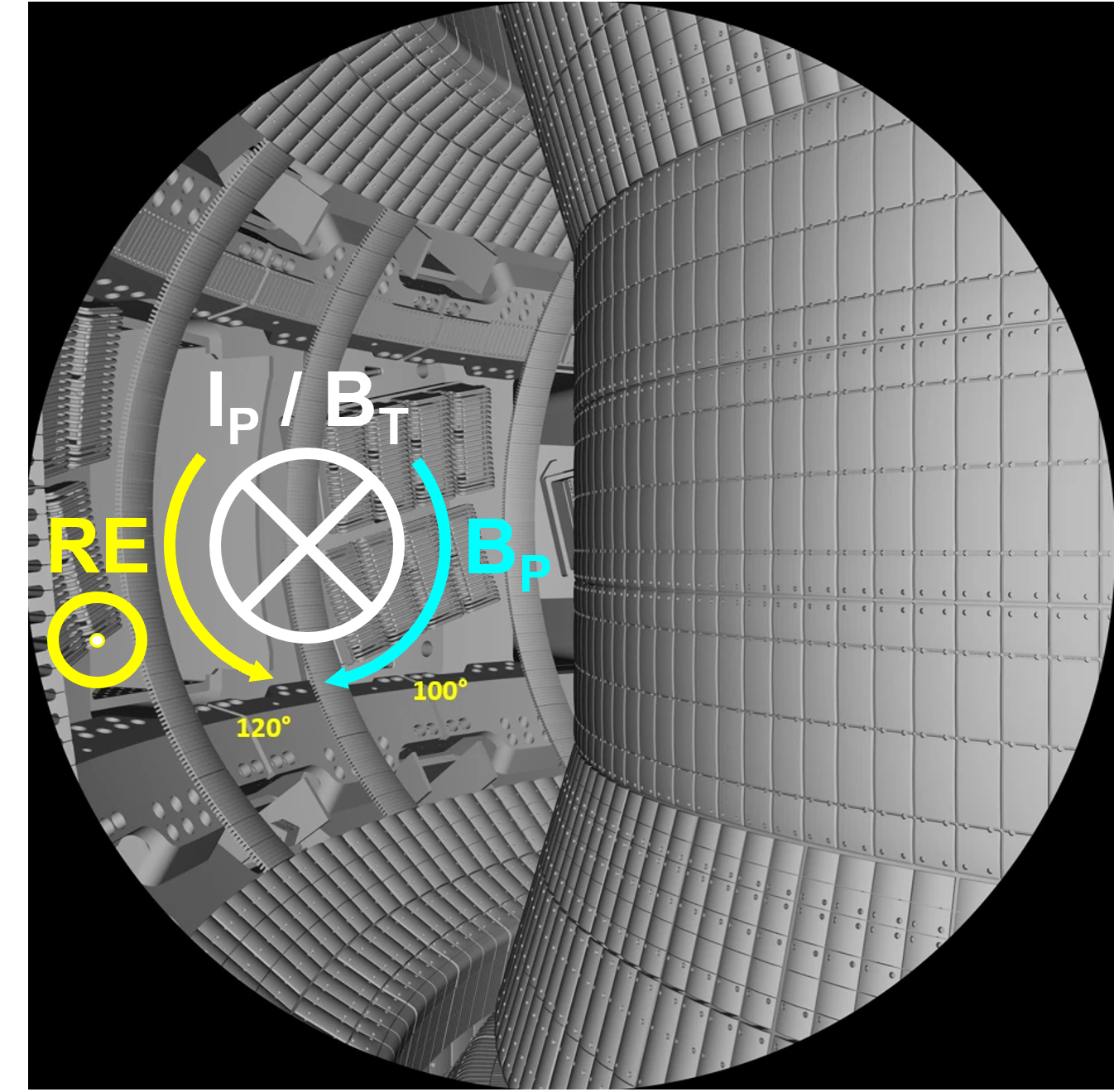}
            \caption{Clockwise (CW)}
            \label{fig:cam-view-cw}
        \end{subfigure}
        \begin{subfigure}{\mywidth}
            \centering
            \includegraphics[height=4cm]{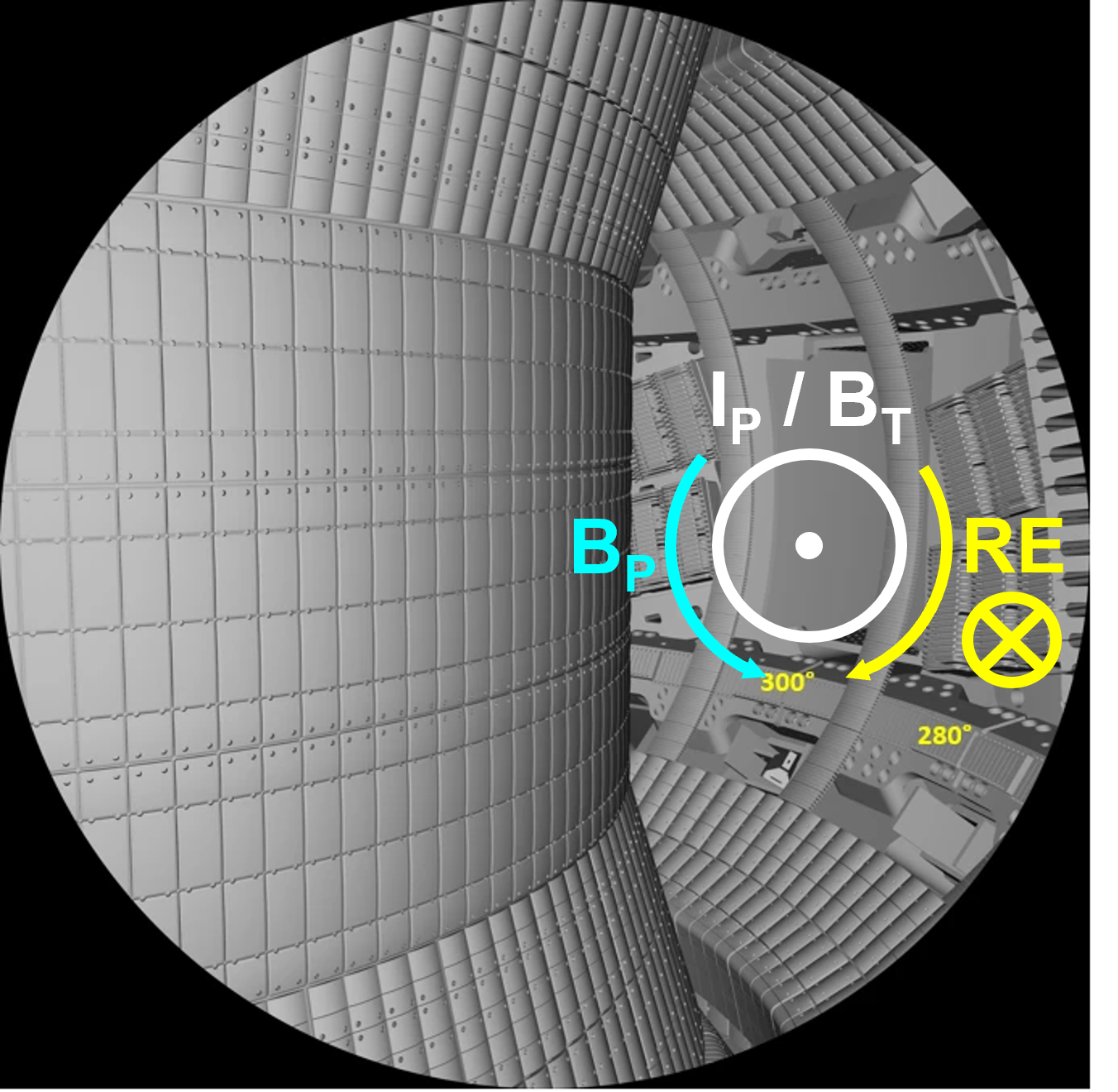}
            \caption{Counter-clockwise}
            \label{fig:cam-view-ccw}
        \end{subfigure}
        %\caption{Proposed (a)~clockwise (CW) and (b)~counterclockwise wide-angle views co-located at the same outboard midplane port in SPARC. The nominal CW direction of the plasma current $\Ip$ and toroidal magnetic field $\Bt$ are shown, along with the poloidal magnetic field $\Bp$ and RE direction. Angles indicate (arbitrary) toroidal port locations.}
        \caption{Proposed \reved{(a)~clockwise (CW) and (b)~counter-clockwise (CCW)} wide-angle views co-located at the same outboard midplane port in SPARC. The nominal CW direction of the plasma current $\Ip$ and toroidal magnetic field $\Bt$ are shown, along with the poloidal magnetic field $\Bp$ and RE direction. Angles indicate toroidal port locations.}
        \label{fig:camera-views}
    \end{figure}

\begin{figure}%[h!]
        \centering
        \includegraphics[width=0.95\columnwidth]{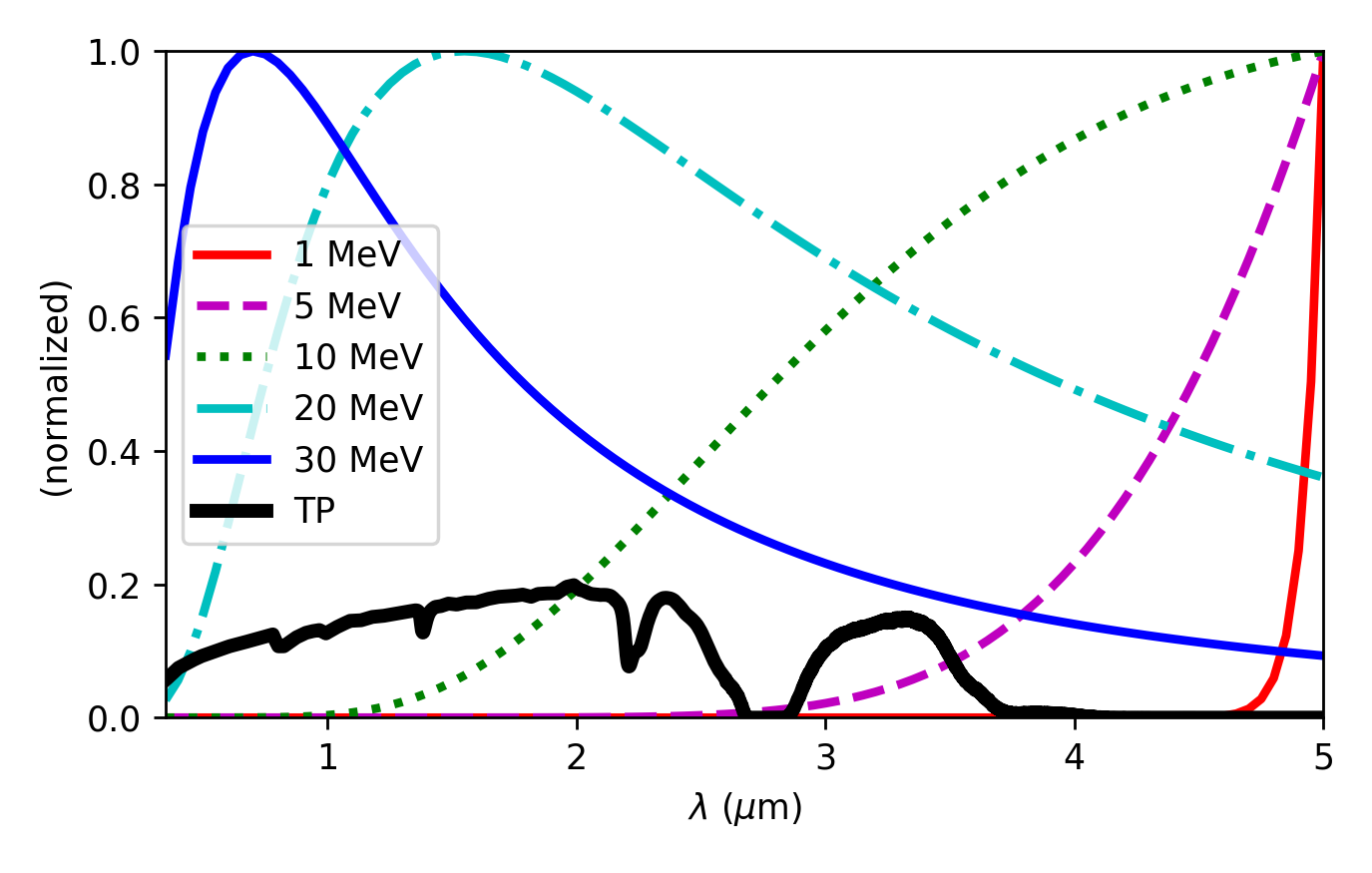}
        \caption{Throughput (TP) of light via proposed in-vessel optics, for wavelengths $\lambda = \SI{0.35 - 5}{\mu m}$, is thick black. Synthetic synchrotron spectra, normalized to their respective maxima, are shown for RE energies $\sim$ $\SI{1}{MeV}$ (solid, at right), $\SI{5}{MeV}$ (dashed), $\SI{10}{MeV}$ (dotted), $\SI{20}{MeV}$ (dot-dashed), and $\SI{30}{MeV}$ (solid, at left). \revone{The pitch angle distribution is uniform over $\thetap = \SI{0-0.1}{rad}$.}}
        \label{fig:spectra_tp}
    \end{figure}

    Preliminary specifications for the CW FOV are given in \cref{tab:cam_params}. In addition, a relay of four mirrors \add{and two vacuum windows}, within a midplane port, are planned to transmit light out of the tokamak vacuum vessel. The expected throughput, i.e. light transmission, is plotted in \cref{fig:spectra_tp} over visible and IR wavelengths \add{of interest}, $\lambda = \SI{350}{nm} - \SI{5}{\mu m}$. At most, about 20\% of light is transmitted, and this importantly does not include the light path from the tokamak to the diagnostic hall or detector efficiencies. There are also regions with low ($<$1\%) throughput, e.g. $\lambda \approx \SI{2.6-2.8}{\mu m}$ and $\lambda > \SI{4.1}{\mu m}$; for this reason, the modeling in \cref{sec:flattop,sec:startup,sec:disruption} only considers wavelengths $\lambda = \SI{350 - 4100}{nm}$. \revone{For cameras specifically, we are currently considering visible \add{($\SI{350-800}{nm}$)}, short-IR \add{($\SI{1.55-1.65}{\mu m}$)}, and mid-IR \add{($\SI{3.45-3.55}{\mu m}$)} wavelength ranges. Note that the IR cameras are filtered to avoid line radiation, but the visible camera is not; however, this could be a possible back-end upgrade in the future.}

    %synthetic images for cameras with sensitivities in the visible \add{($\SI{350-800}{nm}$)}, short-IR \add{($\SI{1.55-1.65}{\mu m}$)} and mid-IR \add{($\SI{3.45-3.55}{\mu m}$)} wavelength ranges
    
    %\revone{We note here that }
    
    \begin{table}%[h!]
        \centering
        \caption{Proposed parameters for the clockwise wide-angle field of view in \cref{fig:cam-view-cw}.}
        \label{tab:cam_params}
        \begin{tabular}{l r}
            \hline
            Major radial position $R$ & $\SI{2.75}{m}$ \\
            Vertical position $Z$ & $\SI{-6}{cm}$ \\
            Tilt (above midplane) &  \revtwo{$+3^\circ$} \\% $\SI{+3}{deg}$ \\
            Full opening angle & \revtwo{$48^\circ$} \\%$\SI{48}{deg}$ \\
            Aperture diameter & $\SI{8}{mm}$ \\
            \hline
        \end{tabular}
    \end{table}
    %\vspace{cm}

    The Synchrotron Orbit-Following Toolkit (SOFT) \cite{Hoppe2018} is used to generate synthetic images and spectra in this work. As a brief overview of the code, an axisymmetric magnetic equilibrium is input, along with an aperture diameter, position, viewing direction, opening angle, and spectral range (see \cref{tab:cam_params}). RE markers are sampled from specified distributions in space, energy, and pitch angle $\thetap$ (with respect to the local magnetic field, but anti-parallel to $\Ip$). These markers follow field lines, emitting synchrotron radiation in a ``hollow cone'' due to their gyromotion. Finally, the spectral radiance is recorded for any light entering the aperture. Note that wavelength-dependent throughput (\cref{fig:spectra_tp}) is applied in post-processing.
\section{Flat-top runaway electrons}\label{sec:flattop}

    %Can only see REs from $r/a \approx 0.3 - 0.9$. Cone model. gamma = 20, so E ~ 10 MeV. pitch angle 0-0.1.

    %The SPARC tokamak is a compact device, with major and minor radii $R_0 = \SI{1.85}{m}$ and $a = \SI{0.57}{m}$, respectively \cite{Creely2020}. Its high on-axis magnetic field strength, $B_0 = \SI{12.2}{T}$, is enabled by high temperature superconducting technology \cite{Hartwig2024}. The main mission of SPARC is to demonstrate fusion breakeven, $Q > 1$, in a magnetic confinement device; the highest power ``Primary Reference Discharge'' (PRD) is expected to achieve $Q \sim 10$.

    Magnetic equilibria for the \revtwo{SPARC} PRD are shown in \cref{fig:startup_eq,fig:flattop_eq} during plasma start-up and flat-top, when $\Ip~=~\SI{0.2}{MA}$ and $\SI{8.7}{MA}$, respectively. 
    %The latter is a double-null magnetic configuration. 
    The steady-state electron density for the PRD is $n_e > \SI{10^{20}}{m^{-3}}$, while the loop voltage required to drive the plasma current is $O(\SI{1}{V})$; therefore, $E/\Ec \sim O(1)$, and no REs are expected during normal operations. However, the flat-top phase can be used as a (pedagogical) reference against which start-up and disruptions can be compared. 

    %The steady-state electron density is quite high for the PRD, $n_e > \SI{10^{20}}{m^{-3}}$, while the loop voltage required to drive the plasma current is low $O(\SI{1}{V})$; therefore, $E/\Ec \sim O(1)$, and no REs are expected during normal operations. However, the flat-top phase can be used as a (pedagogical) reference against which start-up and disruption can be compared. 

    \renewcommand{\mywidth}{0.4\columnwidth}
    \renewcommand{\myheight}{5.2cm}
    \begin{figure}[h!]
        \centering
        \begin{subfigure}{\mywidth}
            \centering
            \includegraphics[height=\myheight]{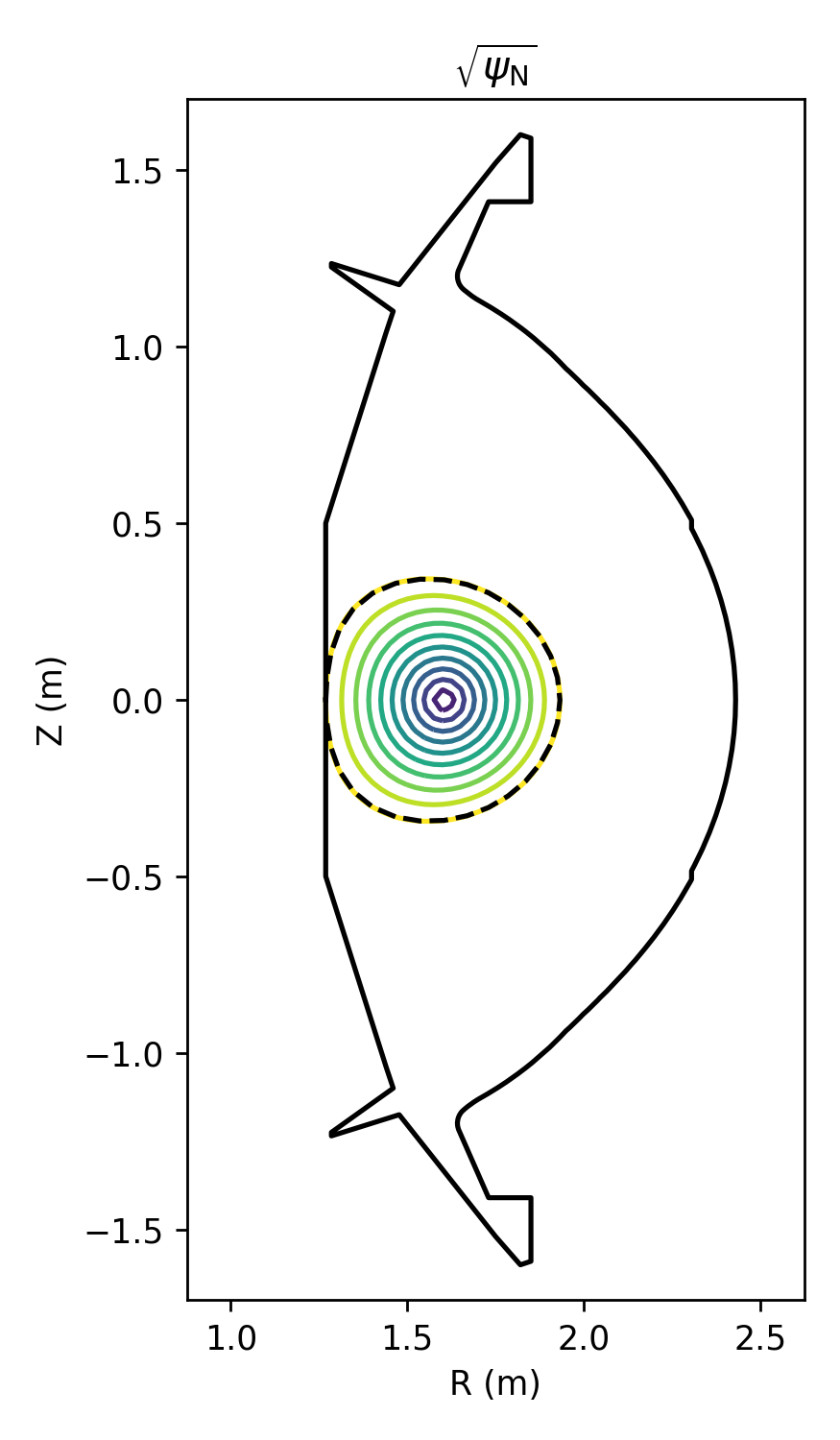}
            \caption{$\Ip = \SI{0.2}{MA}$}
            \label{fig:startup_eq}
        \end{subfigure}
        \begin{subfigure}{\mywidth}
            \centering
            \includegraphics[height=\myheight]{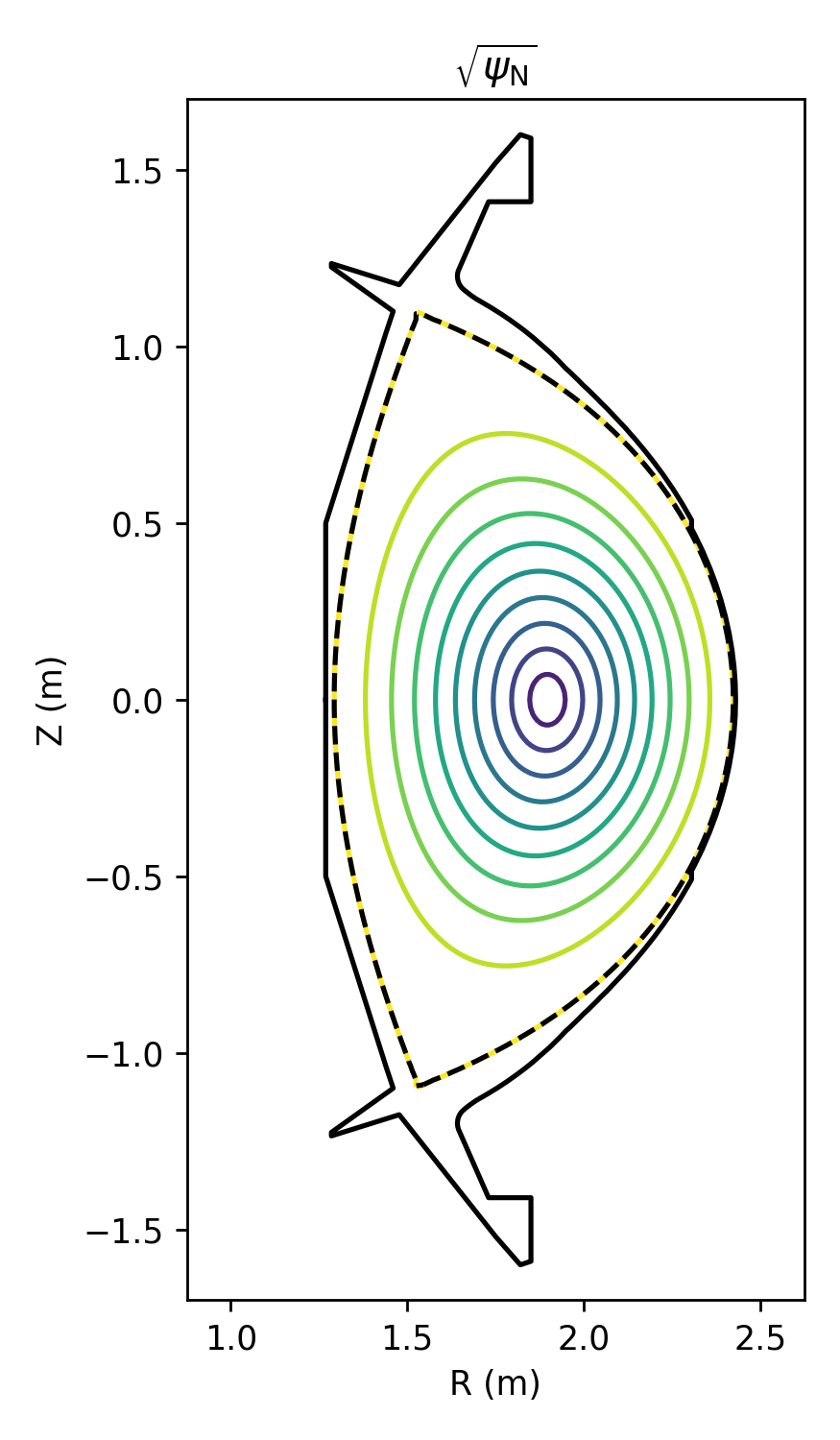}
            \caption{$\Ip = \SI{8.7}{MA}$}
            \label{fig:flattop_eq}
        \end{subfigure}
        \caption{Plasma equilibria for (a)~start-up and (b)~flat-top phases \add{of the PRD}. Contours of the square root of normalized poloidal flux ($\sqrtpsin = 0.1, 0.2,\dots, 1$) are shown along with the last closed flux surface (dashed) and first wall (solid).}
        \label{fig:startup_flattop_eq}
    \end{figure}

    \Cref{fig:flattop_all} demonstrates the capabilities of SOFT for the flat-top equilibrium, from \revtwo{TRANSP simulations \cite{Rodriguez-Fernandez2020}}, in \cref{fig:flattop_eq}. First, REs are populated on individual flux surfaces spanning $r/a~=~{0-1}$ at 0.1 increments. A synthetic synchrotron image \add{(for this unrealistic distribution function)} is shown in \cref{fig:flattop_bands}.
    \revtwo{
        Here (and in \cref{fig:startup_bands,fig:disr_bands}), the synchrotron brightness from each flux surface has been normalized so that all surfaces can be seen on the same (arbitrary) colorscale.
    }
    Again, this is the CW view from \cref{fig:cam-view-cw}, so that the low-field and high-field sides (\add{LFS and HFS,} or outboard and inboard walls) are on the left and right, respectively. We can ``see'' REs from the plasma center to the last closed flux surface, but primarily on the HFS.

    \renewcommand{\mywidth}{0.49\columnwidth}
    \renewcommand{\myheight}{3.25cm}
    \begin{figure}%[h!]
        \centering
        \begin{subfigure}{\mywidth}
            \centering
            \includegraphics[height=\myheight]{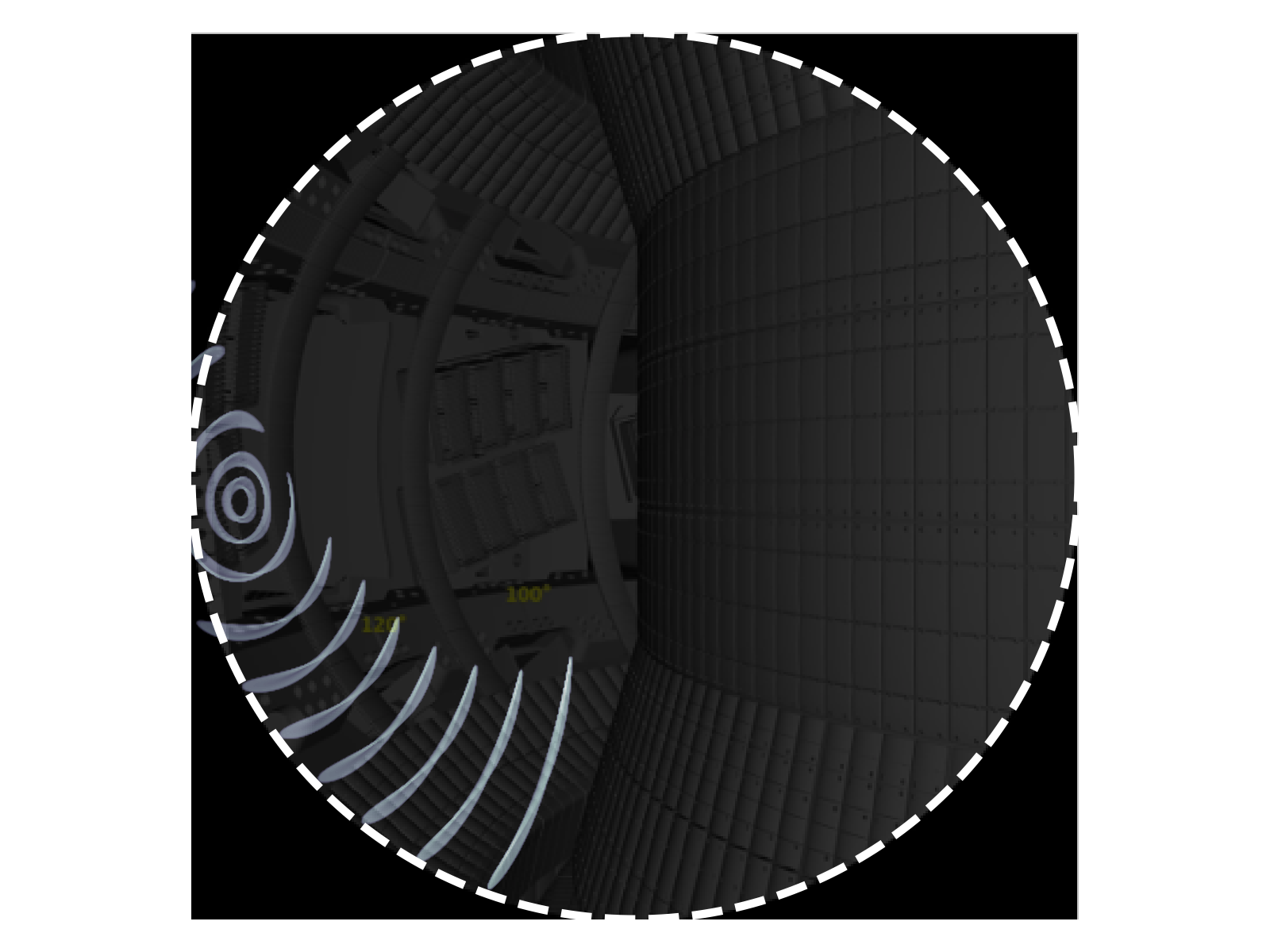}
            \caption{}
            \label{fig:flattop_bands}
        \end{subfigure}
         \begin{subfigure}{\mywidth}
            \centering
            \includegraphics[height=3cm]{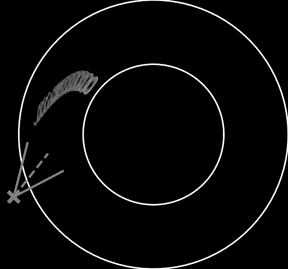}
            \caption{}
            \label{fig:flattop_bands_top}
        \end{subfigure}
        \begin{subfigure}{\mywidth}
            \centering
            \includegraphics[height=\myheight]{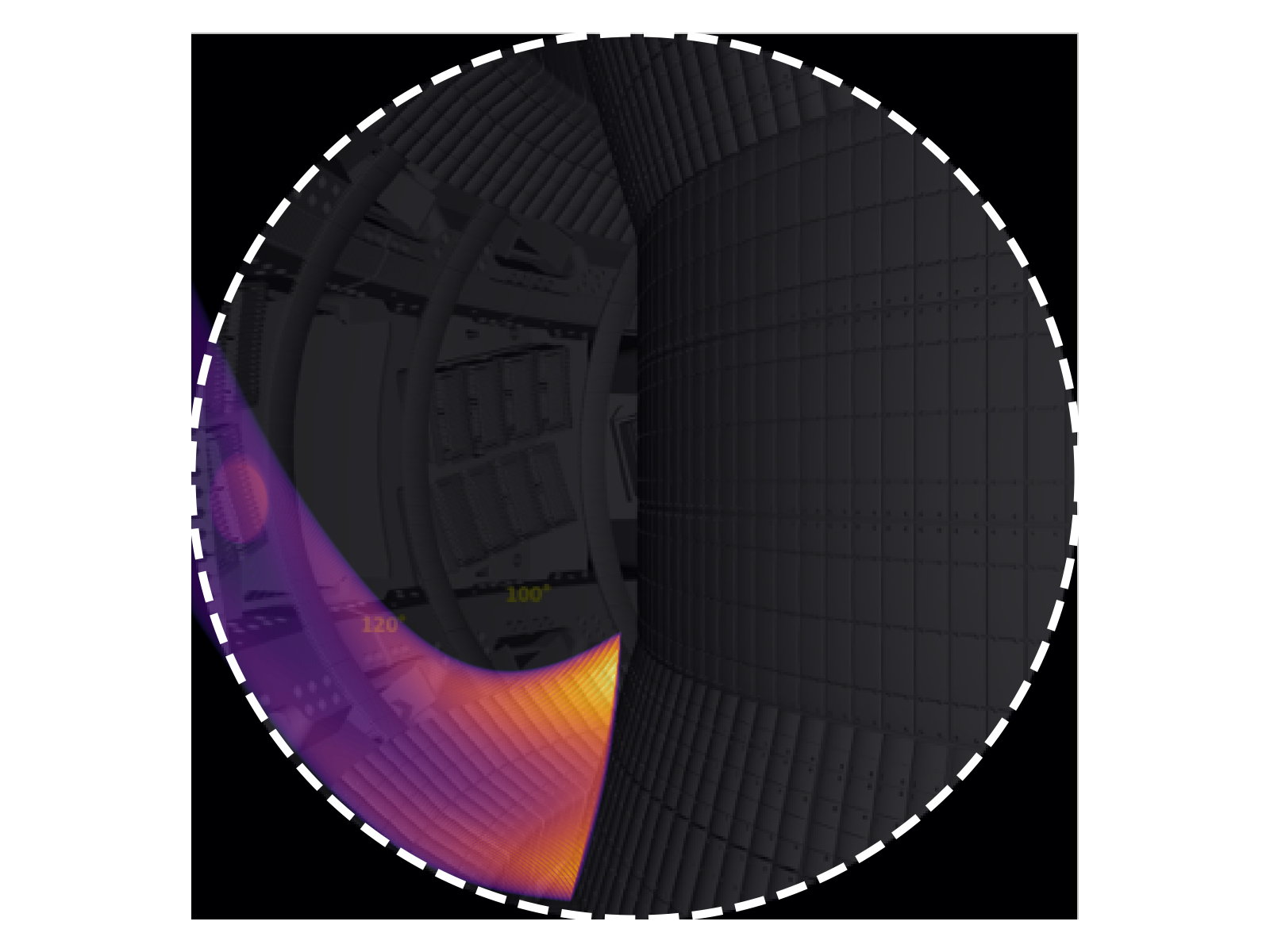}
            \caption{}
            \label{fig:flattop_full}
        \end{subfigure}
        \begin{subfigure}{\mywidth}
            \centering
            \includegraphics[height=3cm]{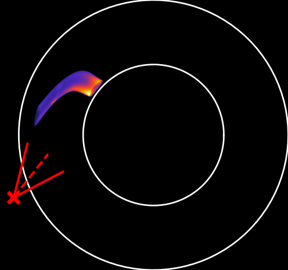}
            \caption{}
            \label{fig:flattop_full_top}
        \end{subfigure}
        \caption{(a,c)~Synthetic camera images for REs populated in $r/a = 0-1$ (a)~at discrete intervals $\mathrm{d}r/a = 0.1$ and (c)~uniformly for the flat-top equilibrium in \cref{fig:flattop_eq}. Colorscale is intensity (a.u.); FOV boundary is dashed. (b,d)~Top-down views of $(R,\phi)$ positions where REs emit the detected synchrotron radiation. The aperture is at the `$\times$' with viewing direction dashed and FOV solid. Inner and outer walls are solid circles.} 
        \label{fig:flattop_all}
    \end{figure}

    %Because synchrotron emission is so highly directional, we mostly see REs as they traverse below the midplane ($Z < 0$), opposite the direction of the poloidal magnetic field $\Bp$ (see \cref{fig:cam-view-cw}), and ``shine'' light toward the aperture. Thus, observed synchrotron emission is strongly dependent on the magnetic geometry. This is confirmed in the ``top-down view'' in \cref{fig:flattop_bands_top}. It is important to note that this is not a synthetic image, but rather a toroidal map of locations where REs are detected;  the grey ``ovals'' indicate the locations where synchrotron light is directed at \add{and detected by} the aperture ($\times$). %Once again, we observe how synchrotron emission is highly anisotropic.

    Because synchrotron emission is so highly directional, we mostly see REs as they traverse below the midplane ($Z < 0$), opposite the direction of the poloidal magnetic field $\Bp$ (see \cref{fig:cam-view-cw}), and ``shine'' light toward the aperture. 
    \revtwo{
        Thus, observed synchrotron emission from a given plasma position is strongly dependent on the local magnetic geometry. 
    }
    This is confirmed in the ``top-down view'' in \cref{fig:flattop_bands_top}. It is important to note that this is not a synthetic image, but rather a toroidal map of locations where REs are detected;  the grey ``ovals'' indicate the locations where synchrotron light is directed at \add{and detected by} the aperture ($\times$). 

    \Cref{fig:flattop_full,fig:flattop_full_top} are reproductions of \cref{fig:flattop_bands,fig:flattop_bands_top}, except that REs are now uniformly distributed throughout the plasma. The RE energy is ${\sim}\SI{10}{MeV}$ and pitch angle distribution is uniform over $\thetap = \SI{0-0.1}{rad}$. The colorscale is false-colored (i.e. not what a camera would truly measure) and instead represents the normalized pixel intensity, which is brightest on the HFS since radiated power scales as $P \propto (p \, \sin{\thetap} \, B)^2 $, with $p$ the RE relativistic momentum. 

    %$P \propto p^2 \sin^2{\thetap} B^2 $

    \revone{
        This pitch angle range is chosen because RE velocity-space distribution functions are often highly elongated in the parallel direction $\vpar$ (opposite $\Ip$), with phase-space density falling off exponentially beyond $\vperp/\vpar > 0.1$; yet the ``brightest'' region of phase space is often localized around $\vperp/\vpar \sim 0.1$ (for example, see Fig.~8 of \cite{Tinguely2018spectra}) due to the $\sin^2{\thetap}$ power scaling. Although not shown here, larger pitch angles make the image ``wider'' or ``taller'', as one might expect, allowing some part of the distribution function to be assessed in this way. 
    } 
\section{Start-up runaway electrons}\label{sec:startup}
    %\newpage
    %HXR \cite{Panontin2024}

    While hard x-ray detectors will be the primary monitor of start-up REs (from thin and thick-target bremsstrahlung) in SPARC \cite{Panontin2024}, it is of interest to explore whether synchrotron emission can be a complementary diagnostic measurement. An inner-wall-limited start-up equilibrium, generated with the \add{free-boundary tokamak equilibrium} \revone{(FBT)} code \cite{Hofmann1988}, is seen in \cref{fig:startup_eq}. With a (planned) plasma current ramp rate of ${\sim}\SI{1}{MA/s}$, the equilibrium at $t = \SI{0.2}{s}, \Ip = \SI{0.2}{MA}$, is somewhat ``far along'' during start-up. Still, the high-safety-factor, small-bore plasma is representative of the magnetic geometry in which start-up REs could form.
    \revtwo{
        Moreover, because REs can maximally carry the full plasma current, a time resolution of order ${\sim}\SI{10}{ms}$, i.e. RE current increments ${\sim}\SI{10}{kA}$, should be sufficient to ``catch'' start-up RE formation.  
    }

    %Like \cref{fig:flattop_bands}, \cref{fig:startup_bands} shows REs populated on individual flux surfaces; however, in contrast to the flat-top equilibrium (\cref{fig:flattop_eq}), synchrotron emission is detected from start-up REs throughout almost the entire plasma. This is simply due to the plasma's small cross-section and inner-wall localization. Thus, the CW view (\cref{fig:cam-view-cw}) should adequately capture synchrotron radiation from any start-up RE during early plasma formation.   

    Like \cref{fig:flattop_bands}, \cref{fig:startup_bands} shows REs populated on individual flux surfaces; however, in contrast to the flat-top equilibrium (\cref{fig:flattop_eq}), \revtwo{there is improved coverage of the poloidal cross-section, including both above and below the midplane.}
    This is simply due to the plasma's small cross-section and inner-wall localization. Thus, the CW view (\cref{fig:cam-view-cw}) should adequately capture synchrotron radiation from any start-up RE during early plasma formation.

    \renewcommand{\mywidth}{0.49\columnwidth}
    \renewcommand{\myheight}{3.25cm}
    \begin{figure}%[h!]
        \centering
        \begin{subfigure}{\mywidth}
            \centering
            \includegraphics[height=\myheight]{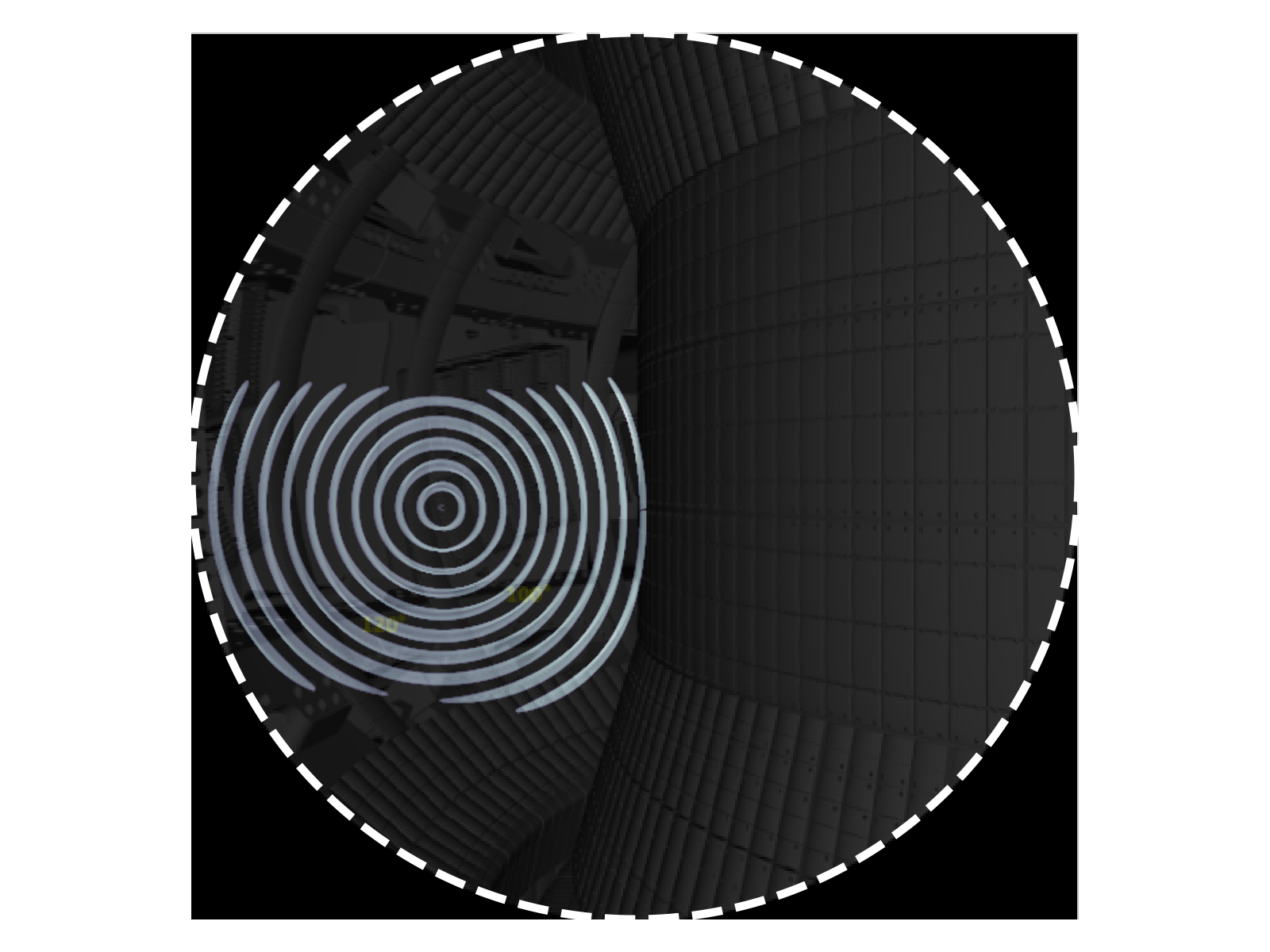}
            \caption{$r/a = 0-1$, $\mathrm{d}r/a = 0.1$}
            \label{fig:startup_bands}
        \end{subfigure}
        \begin{subfigure}{\mywidth}
            \centering
            \includegraphics[height=\myheight]{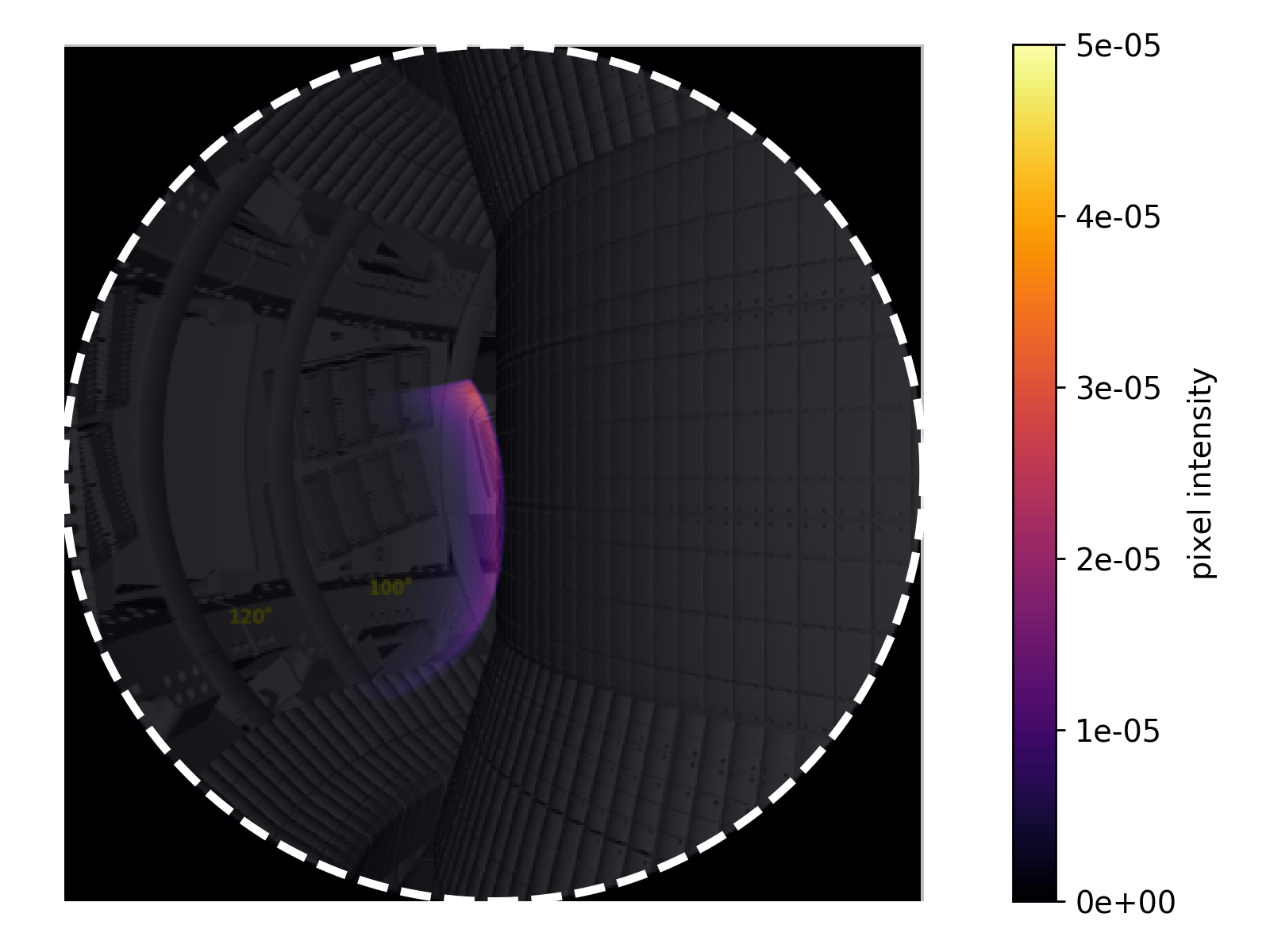}
            \caption{$\lambda = \SI{350-800}{nm}$}
            \label{fig:startup_vis}
        \end{subfigure}
        \begin{subfigure}{\mywidth}
            \centering
            \includegraphics[height=\myheight]{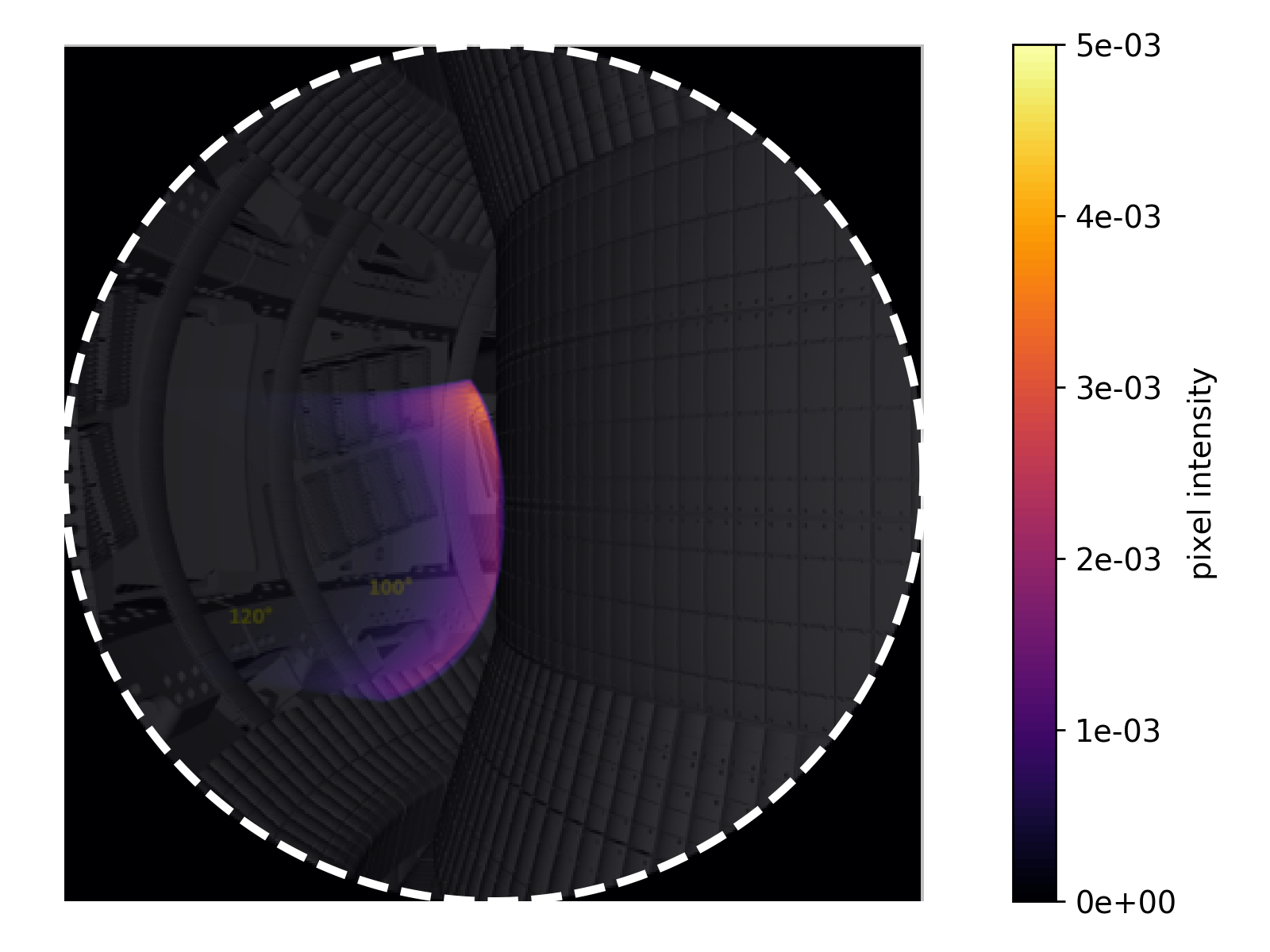}
            \caption{$\lambda = \SI{1.55-1.65}{\mu m}$}
            \label{fig:startup_swir}
        \end{subfigure}
        \begin{subfigure}{\mywidth}
            \centering
            \includegraphics[height=\myheight]{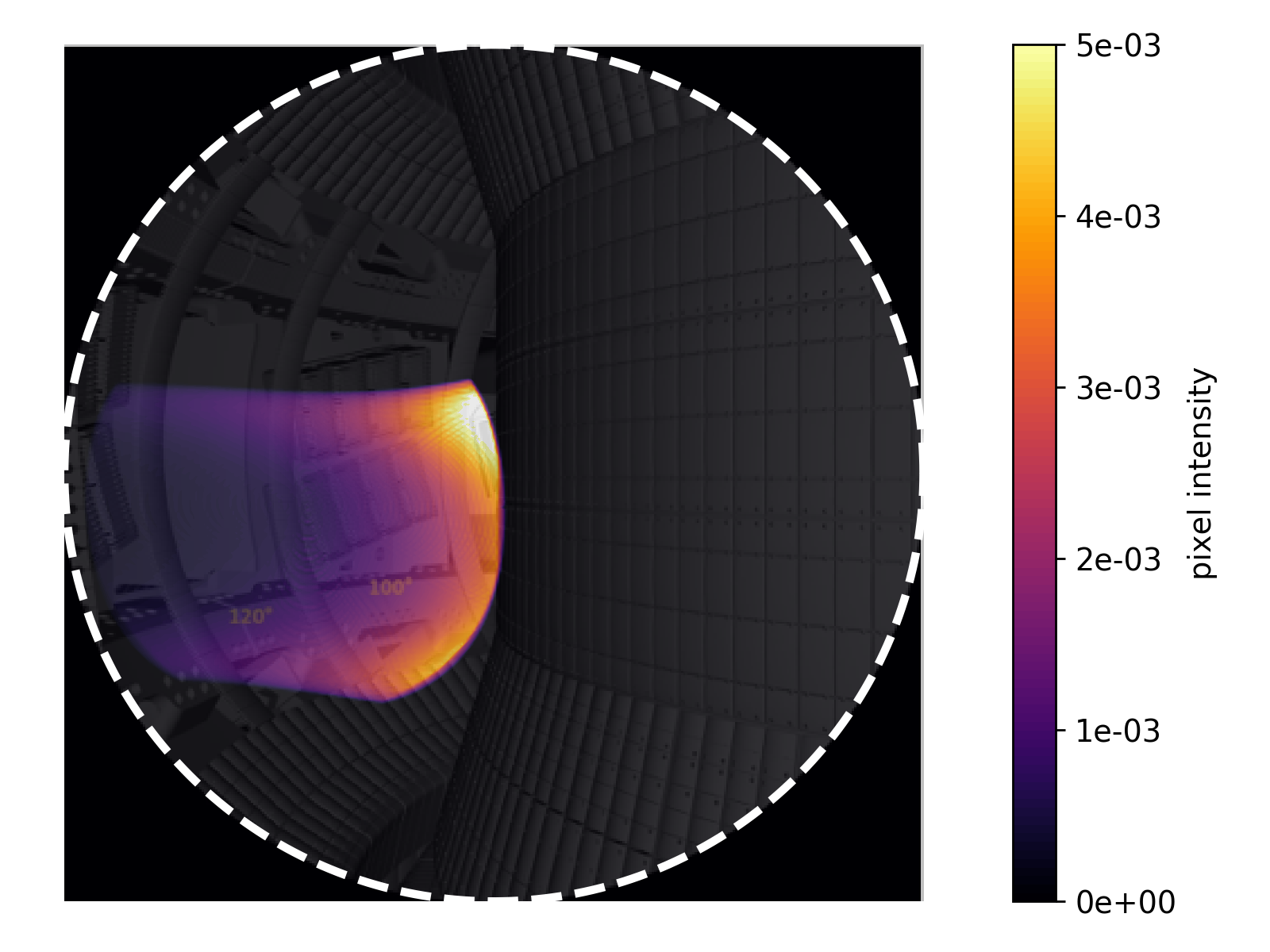}
            \caption{$\lambda = \SI{3.45-3.55}{\mu m}$}
            \label{fig:startup_mwir}
        \end{subfigure}
        \caption{Synthetic camera images for REs populated (a)~on discrete radial bands and (b)-(d)~uniformly for the start-up equilibrium in \cref{fig:startup_eq}; throughput from \cref{fig:spectra_tp} is applied when integrating over each spectral range. Colorscale is pixel intensity, normalized to the maximum integrated over $\lambda = \SI{0.35-4.1}{\mu m}$. FOV boundary is dashed. RE energy is ${\sim}\SI{10}{MeV}$, and pitch angles are uniform over $\thetap = \SI{0-0.1}{rad}$.} 
        \label{fig:startup_bands_spec}
    \end{figure}

    %Adequate signal level is another issue, which depends on the RE distribution function (energies, pitch angles, and number) as well as optical throughput and detector efficiency. The loop voltage during the $\Ip$ ramp-up is $O(\SI{10}{V})$, meaning that REs can maximally gain ${\sim}\SI{250}{keV/ms}$; e.g. $\SI{2.5}{MeV}$ in $\SI{10}{ms}$ or $\SI{25}{MeV}$ in $\SI{100}{ms}$. The former seems realistic for start-up REs, while the latter \reved{could} be limited by such mechanisms as bremsstrahlung and synchrotron radiation.
    %\revtwo{At this stage, the brightness of synchrotron emission in SPARC is largely uncertain, and we will leave its estimate for future work.}
    %\add{At this stage, the brightness of synchrotron emission in SPARC is largely uncertain; however, for reference, spectral radiance values of order $\SI{0.1}{\mu W/mm^2/sr/nm}$ were measured by visible spectrometers in Alcator C-Mod, from multi-MeV REs at $B_0 = \SI{7.8}{T}$ \cite{Tinguely2018spectra}.}

    \revtwo{
        The loop voltage during the $\Ip$ ramp-up is $O(\SI{10}{V})$, meaning that REs can maximally gain ${\sim}\SI{250}{keV/ms}$; e.g. $\SI{2.5}{MeV}$ in $\SI{10}{ms}$ or $\SI{25}{MeV}$ in $\SI{100}{ms}$. The former seems realistic for start-up REs, while the latter could be limited by such mechanisms as bremsstrahlung and synchrotron radiation or even confinement loss due to large drift orbits.
    }
    \Cref{fig:spectra_tp} shows synthetic synchrotron spectra for RE energies spanning $\SI{1-30}{MeV}$. The spectral peak shifts toward shorter wavelengths (from IR to visible) as RE energy increases. What is more, the expected optical throughput is so low from $\lambda = \SI{4-5}{\mu m}$ that REs with energies ${<}\SI{5}{MeV}$ would be very difficult to detect unless very high in number. Ultimately, a code like STREAM \add{(STart-up Runaway Electron Analysis Model)} \cite{Hoppe2022} should be used to self-consistently evolve start-up RE generation; for this analysis, we consider a simple flat spatial profile of REs with energy ${\sim}\SI{10}{MeV}$ and pitch angles uniform over $\thetap = \SI{0-0.1}{rad}$, \add{i.e. REs mostly streaming along field lines}.

    \Cref{fig:startup_vis,fig:startup_swir,fig:startup_mwir} show synthetic images for cameras with sensitivities in the visible \add{($\SI{350-800}{nm}$)}, short-IR \add{($\SI{1.55-1.65}{\mu m}$)} and mid-IR \add{($\SI{3.45-3.55}{\mu m}$)} wavelength ranges, respectively. These include the effect of optical throughput from \cref{fig:spectra_tp}. Furthermore, the pixel intensity, though arbitrary in absolute magnitude, is consistent across images; specifically, they are normalized to the maximum pixel intensity when integrated over the entire range $\lambda = \SI{0.35-4.1}{\mu m}$. As expected from the synchrotron spectrum for ${\sim}\SI{10}{MeV}$ REs in \cref{fig:spectra_tp}, 
    \revtwo{
        the mid-IR image in \cref{fig:startup_mwir} is ``brightest'' while the visible image in \cref{fig:startup_vis} is ``dimmest'' 
    }
    (by two orders of magnitude). This suggests that mid (or short) wavelength IR cameras are best suited to detect start-up REs, although only for RE energies ${>}\SI{5}{MeV}$.
    
    %As expected from the synchrotron spectrum for ${\sim}\SI{10}{MeV}$ REs in \cref{fig:spectra_tp}, \cref{fig:startup_mwir} is ``brightest'' while \cref{fig:startup_vis} is ``dimmest'' (by two orders of magnitude). This suggests that mid (or short) wavelength IR cameras are best suited to detect start-up REs, although only for RE energies ${>}\SI{5}{MeV}$.

    \revtwo{
        Adequate signal level is another issue, which depends on the RE distribution function (energies, pitch angles, and number) as well as optical throughput and detector efficiency. At this stage, the absolute brightness of synchrotron emission in SPARC is largely uncertain, and we will leave its estimate for future work.
    }

\section{Disruption runaway electrons}\label{sec:disruption}

    %M3D-C1 \cite{Jardin2012}

    %With $\Ip = \SI{8.7}{MA}$ and electron temperatures $T_e \sim \SI{20}{keV}$, there is a huge potential for RE generation during disruptions in SPARC \cite{Sweeney2020}. In fact, unmitigated RE beams might reach currents of ${\sim}\SI{5}{MA}$ with individual RE energies ${>}\SI{10}{MeV}$ \cite{Tinguely2021REMC}. With the fastest expected current quench (CQ) duration of ${\sim}\SI{3.2}{ms}$, the lifetime of RE beams may be of similar order, e.g. $O(\SI{10}{ms})$. This means that frame rates ${>}\SI{100}{fps}$ are needed to properly diagnose disruption RE physics.

    With $\Ip = \SI{8.7}{MA}$ and electron temperatures $T_e \sim \SI{20}{keV}$, there is a huge potential for RE generation during disruptions in SPARC \cite{Sweeney2020}. In fact, unmitigated RE beams might reach currents of ${\sim}\SI{5}{MA}$ with \revone{average} RE energies ${\sim}\SI{10}{MeV}$ \cite{Tinguely2021REMC} \revone{and maximum, synchrotron-limited energies of even hundreds of MeV \cite{Decker2016}; however, note that this upper limit does not consider all power loss mechanisms (e.g. bremsstrahlung) or confinement loss (e.g. due to drift orbits)}. With the fastest expected current quench (CQ) duration of ${\sim}\SI{3.2}{ms}$, the lifetime of RE beams may be of similar order, e.g. $O(\SI{10}{ms})$. This means that frame rates ${>}\SI{100}{fps}$ are needed to properly diagnose disruption RE physics.

    The nonlinear MHD code M3D-C1 \cite{Jardin2012} is used to model a 2D ``cold'' Vertical Displacement Event (VDE) during a disruption, i.e. after the thermal quench has occurred. (No REs are included in the simulation.) The vertical motion and CQ lasts ${\sim}\SI{10}{ms}$. Magnetic \add{geometries} are shown for three vertical positions in \cref{fig:VDE_eq}. Corresponding synthetic synchrotron images are shown in \cref{fig:disr_bands}, where again REs are populated on distinct radial bands. \Cref{fig:disr_bands_03} is very similar to \cref{fig:flattop_bands}, which makes sense since the magnetic \add{flux surfaces} in \cref{fig:VDE-03cm,fig:flattop_eq} are comparable. As the plasma \add{axis} moves downward, from $Z_0 \approx \SI{-3}{cm}$ to $\SI{-38}{cm}$, the aperture begins to collect light from the ``top'' of the plasma. Luckily, even with a vertical drift of nearly $\SI{40}{cm}$, REs are still visible from the plasma core to edge. However, at $Z_0 \approx \SI{-76}{cm}$, only synchrotron light from the outermost flux surfaces ($r/a \approx 0.8-1$) is barely captured. 

    \renewcommand{\mywidth}{0.3\columnwidth}
    \renewcommand{\myheight}{4.5cm}
    \begin{figure}%[h!]
        \centering
        \begin{subfigure}{\mywidth}
            \includegraphics[height=\myheight]{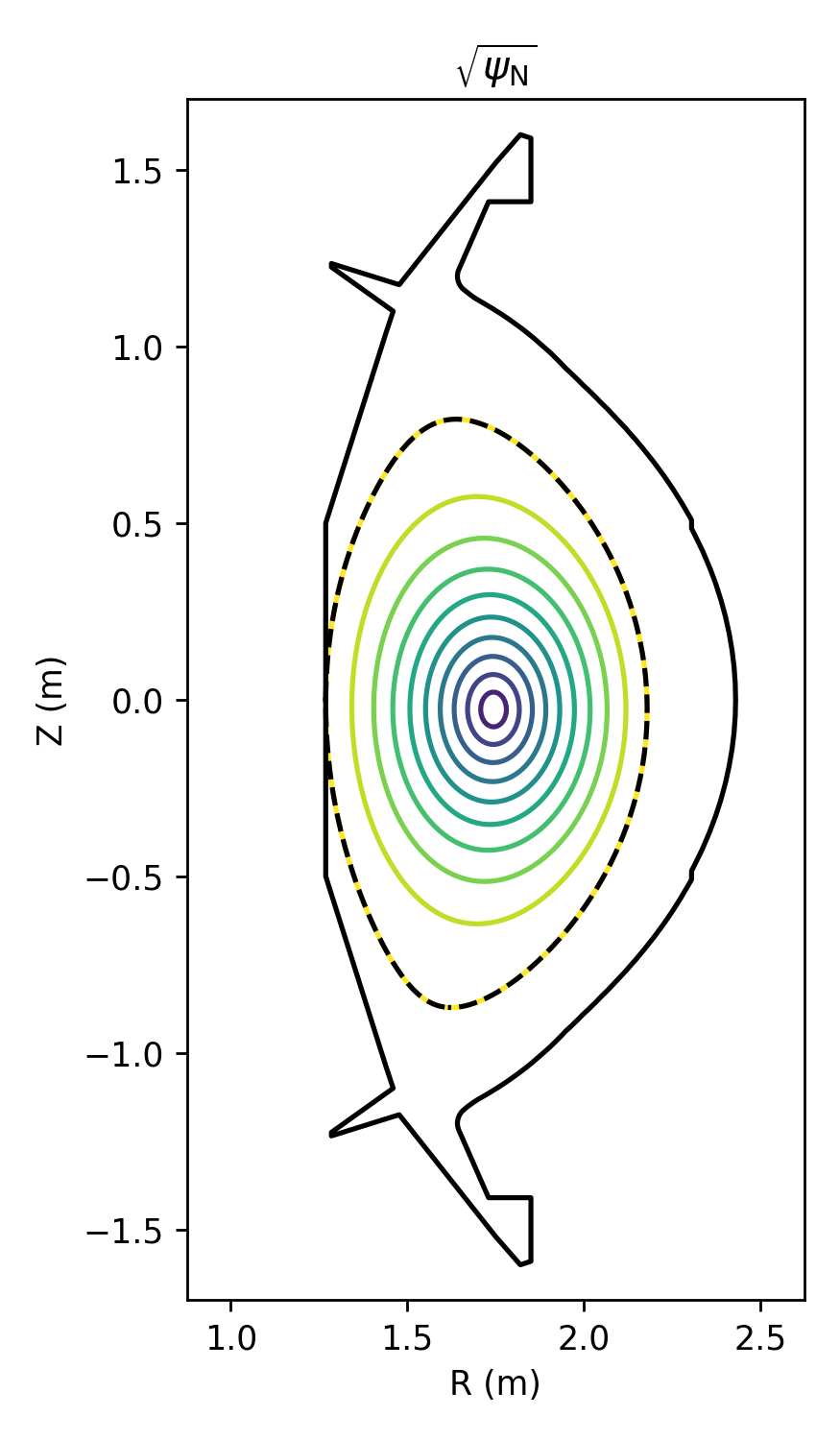}
            \caption{$Z_0\approx\SI{-3}{cm}$}
            \label{fig:VDE-03cm}
        \end{subfigure}
        \begin{subfigure}{\mywidth}
            \includegraphics[height=\myheight]{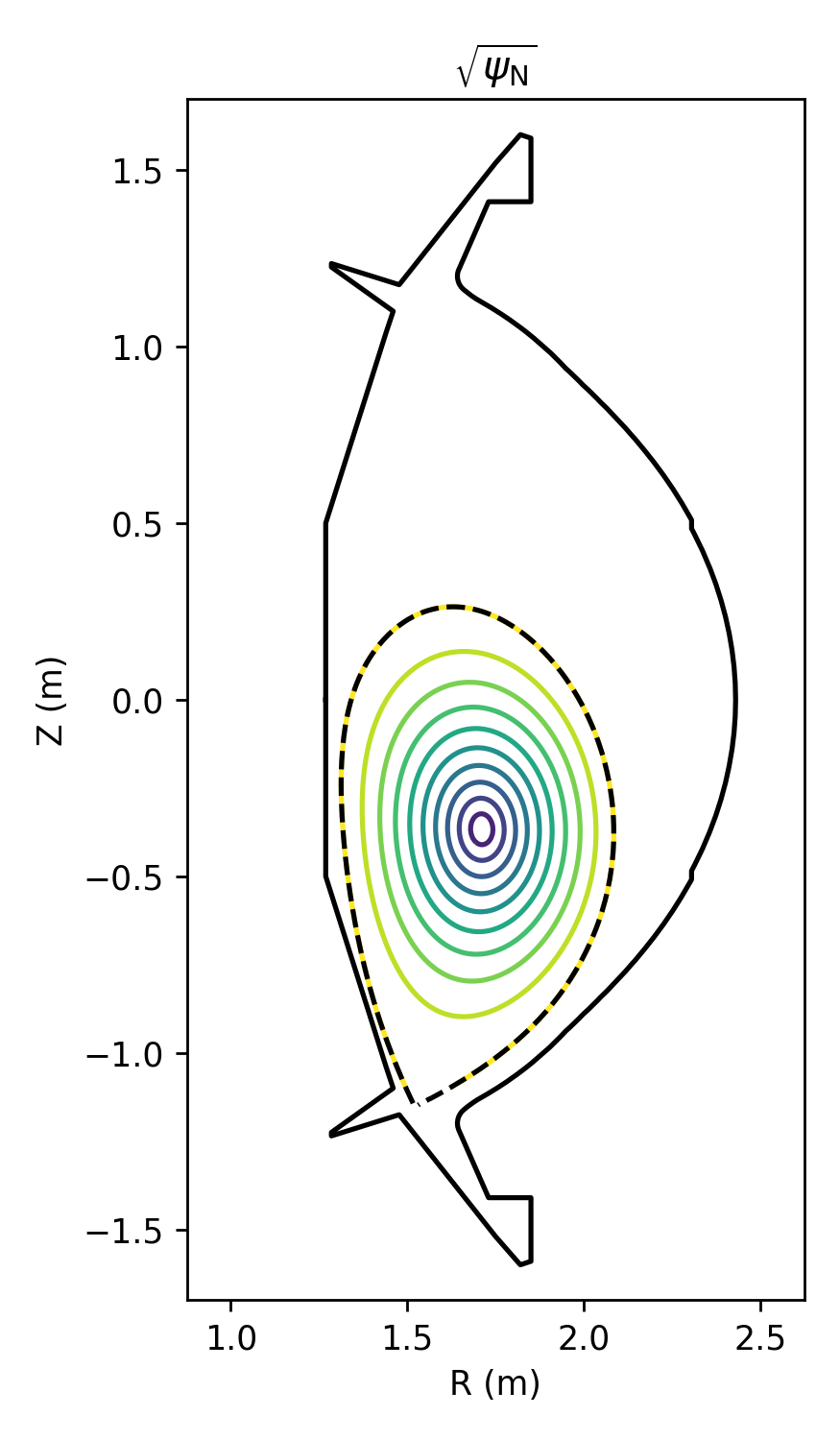}
            \caption{$Z_0\approx\SI{-38}{cm}$}
            \label{fig:VDE-38cm}
        \end{subfigure}
        \begin{subfigure}{\mywidth}
            \includegraphics[height=\myheight]{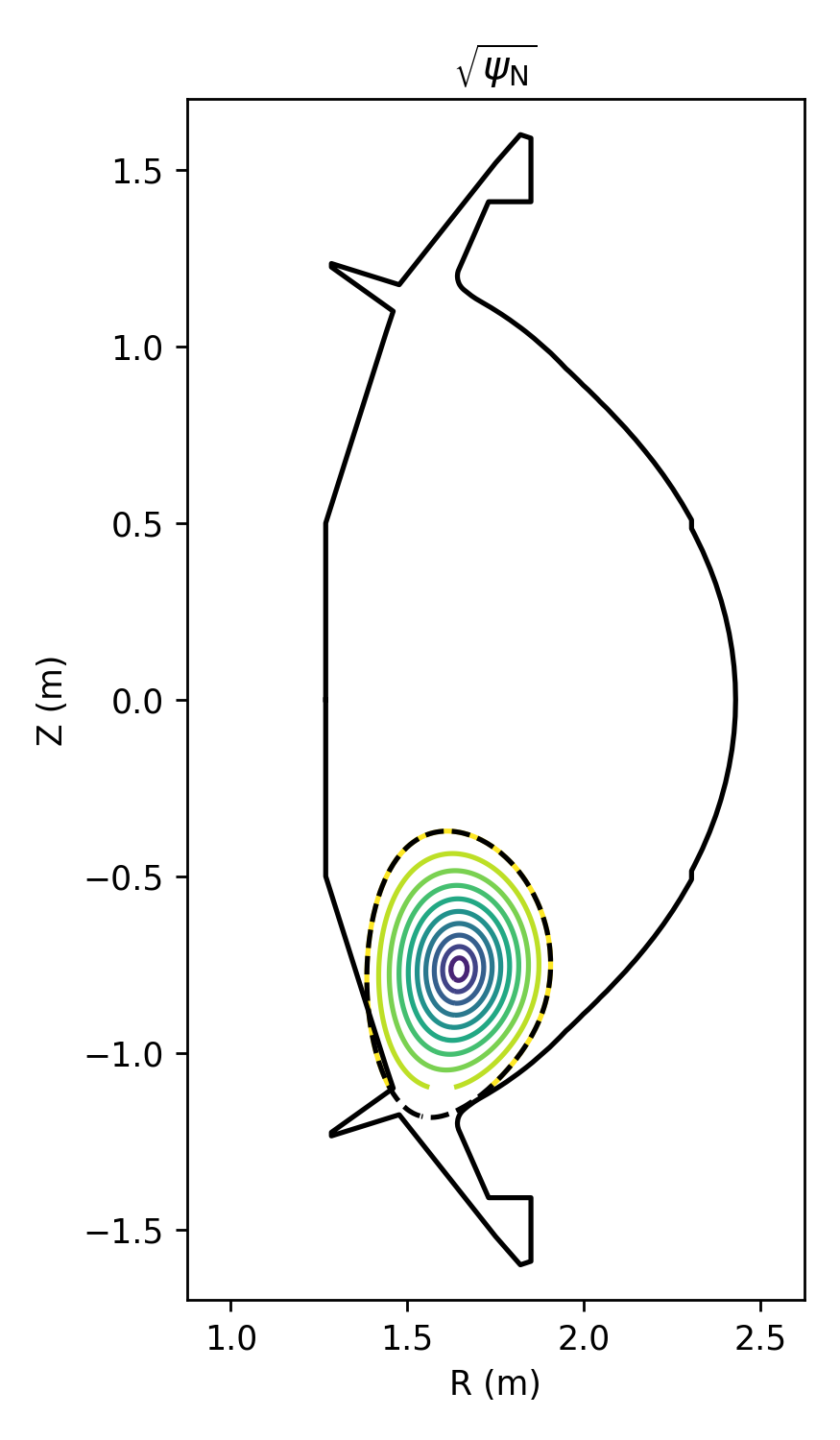}
            \caption{$Z_0\approx\SI{-76}{cm}$}
            \label{fig:VDE-76cm}
        \end{subfigure}
        %\caption{Plasma equilibria during a Vertical Displacement Event at $Z \approx$ (a)~$\SI{-3}{cm}$, (b)~$\SI{-38}{cm}$, and (c)~$\SI{-76}{cm}$.  Contours of the square root of normalized poloidal flux $\sqrtpsin$ are shown along with the last closed flux surface (dashed) and first wall.}
        \caption{Plasma \add{magnetic geometries} during a disruption Vertical Displacement Event \rsiadd{at $Z_0 \approx$ (a)~$\SI{-3}{cm}$, (b)~$\SI{-38}{cm}$, and (c)~$\SI{-76}{cm}$}. Contours of the square root of normalized poloidal flux  ($\sqrtpsin = 0.1, 0.2,\dots, 1$) are shown along with the last closed flux surface (dashed) and first wall (solid).}
        \label{fig:VDE_eq}
    \end{figure}

    %\renewcommand{\mywidth}{0.23\textwidth}
    %\renewcommand{\myheight}{7cm}
    %\begin{figure}[h!]
    %    \centering
    %    \begin{subfigure}{\mywidth}
    %        \centering
    %        \includegraphics[height=4cm]{figs/disruption_radial_bands_Z-03cm.png}
    %        \caption{$Z\approx\SI{-3}{cm}$}
    %        \label{fig:disr_bands_03}
    %    \end{subfigure}
    %    \begin{subfigure}{\mywidth}
    %        \centering
    %        \includegraphics[height=4cm]{figs/disruption_radial_bands_Z-38cm.png}
    %        \caption{$Z\approx\SI{-38}{cm}$}
    %        \label{fig:disr_bands_38}
    %    \end{subfigure}
    %    \begin{subfigure}{\mywidth}
    %        \centering
    %        \includegraphics[height=4cm]{figs/disruption_radial_bands_Z-76cm.png}
    %        \caption{$Z\approx\SI{-76}{cm}$}
    %        \label{fig:disr_bands_76}
    %    \end{subfigure}
    %    \begin{subfigure}{\mywidth}
    %        \centering
    %        \includegraphics[height=4cm]{figs/disruption_spectra.png}
    %        \caption{}
    %        \label{fig:disr_spec}
    %    \end{subfigure}
    %    \caption{(a)-(c)~Synthetic camera image (greyscale) for REs populated on discrete radial bands $r/a \in [0,1]$ at intervals $\mathrm{d}r/a = 0.1$ for the disruption VDE equilibria in \cref{fig:VDE-03cm,fig:VDE-38cm,fig:VDE-76cm}. FOV boundary is dashed. (b)~Normalized synthetic synchrotron spectra for RE energies ${\sim}\SI{10}{MeV}$ (solid), $\SI{20}{MeV}$ (dashed), and $\SI{30}{MeV}$ (dot-dashed).} 
    %    \label{fig:startup_bands_spec}
    %\end{figure}

    \renewcommand{\mywidth}{0.32\columnwidth}
    \renewcommand{\myheight}{2.5cm}
    \begin{figure}%[h!]
        \centering
        \begin{subfigure}{\mywidth}
            \centering
            \includegraphics[height=\myheight]{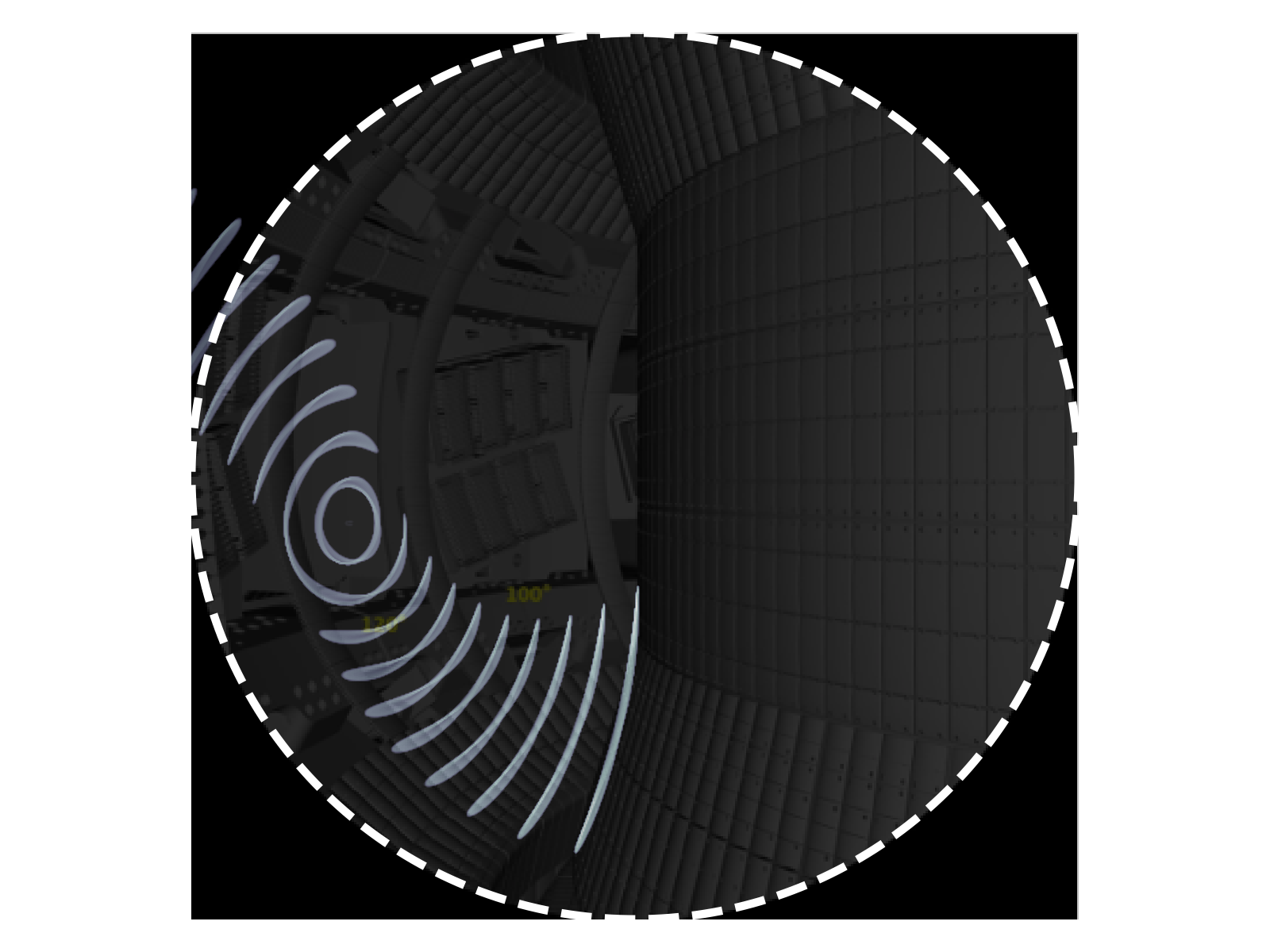}
            \caption{$Z_0\approx\SI{-3}{cm}$}
            \label{fig:disr_bands_03}
        \end{subfigure}
        \begin{subfigure}{\mywidth}
            \centering
            \includegraphics[height=\myheight]{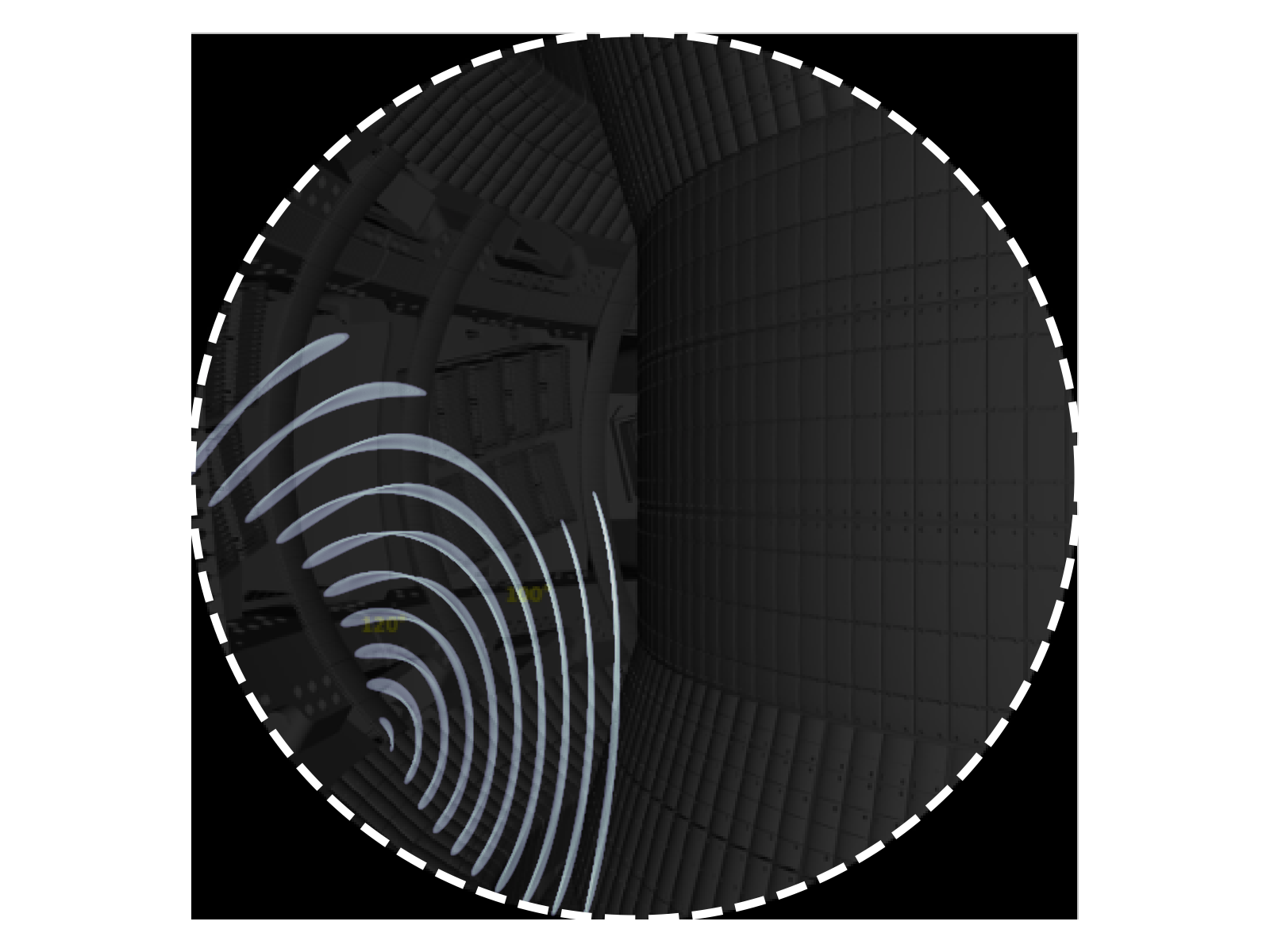}
            \caption{$Z_0\approx\SI{-38}{cm}$}
            \label{fig:disr_bands_38}
        \end{subfigure}
        \begin{subfigure}{\mywidth}
            \centering
            \includegraphics[height=\myheight]{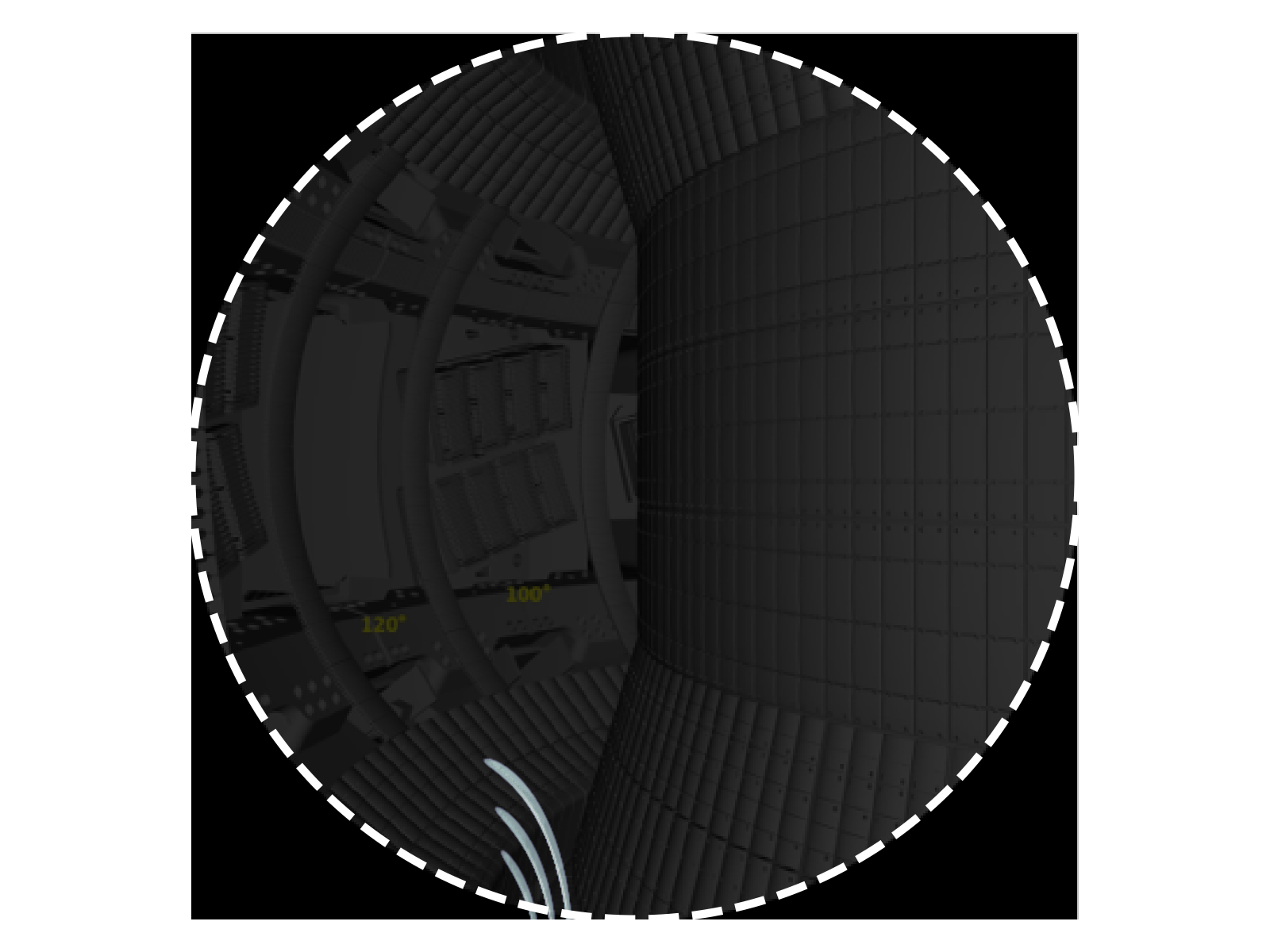}
            \caption{$Z_0\approx\SI{-76}{cm}$}
            \label{fig:disr_bands_76}
        \end{subfigure}
        \caption{(a)-(c)~Synthetic camera image (greyscale) for REs populated on discrete radial bands ($r/a = 0.1, 0.2,\dots, 1$) for the disruption VDE \add{plasmas} in \cref{fig:VDE-03cm,fig:VDE-38cm,fig:VDE-76cm}. FOV boundary is dashed.} 
        \label{fig:disr_bands}
    \end{figure}

    Finally, a similar analysis is carried out as in \cref{sec:startup}, using the VDE \add{magnetic geometry} in \cref{fig:VDE-03cm}. Synthetic images are produced for REs with pitch angles (again) uniformly sampled over $\thetap = \SI{0-0.1}{rad}$, but now with energy ${\sim}\SI{30}{MeV}$. Such high energies are not unphysical, especially with disruption-induced loop voltages ${>}\SI{1}{kV}$. A Gaussian radial profile, in \cref{fig:gaussian}, is chosen as there is some experimental evidence to support such shapes, e.g. in \cite{Tinguely2018}. However, future work will utilize spatial and kinetic distributions from codes like DREAM \add{(Disruption Runaway Electron Analysis Model)} \cite{Hoppe2021DREAM}. 

    \renewcommand{\mywidth}{0.49\columnwidth}
    \renewcommand{\myheight}{3.25cm}
    \begin{figure}%[h!]
        \centering
        \begin{subfigure}{\mywidth}
            \centering
            \includegraphics[height=\myheight]{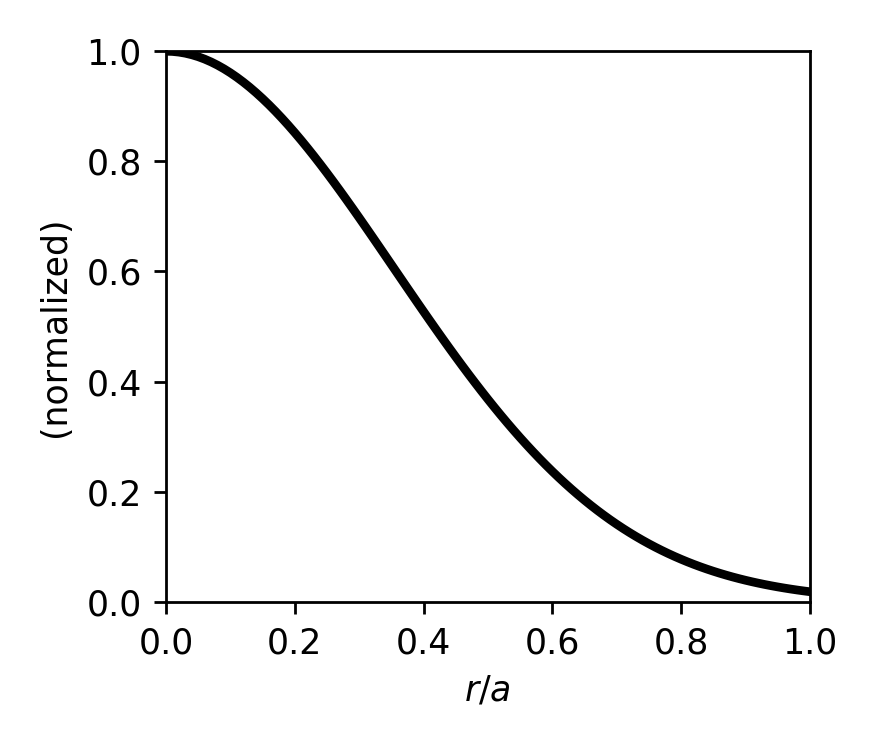}
            \caption{$\propto\exp[-(2r/a)^2]$}
            \label{fig:gaussian}
        \end{subfigure}
        \begin{subfigure}{\mywidth}
            \centering
            \includegraphics[height=\myheight]{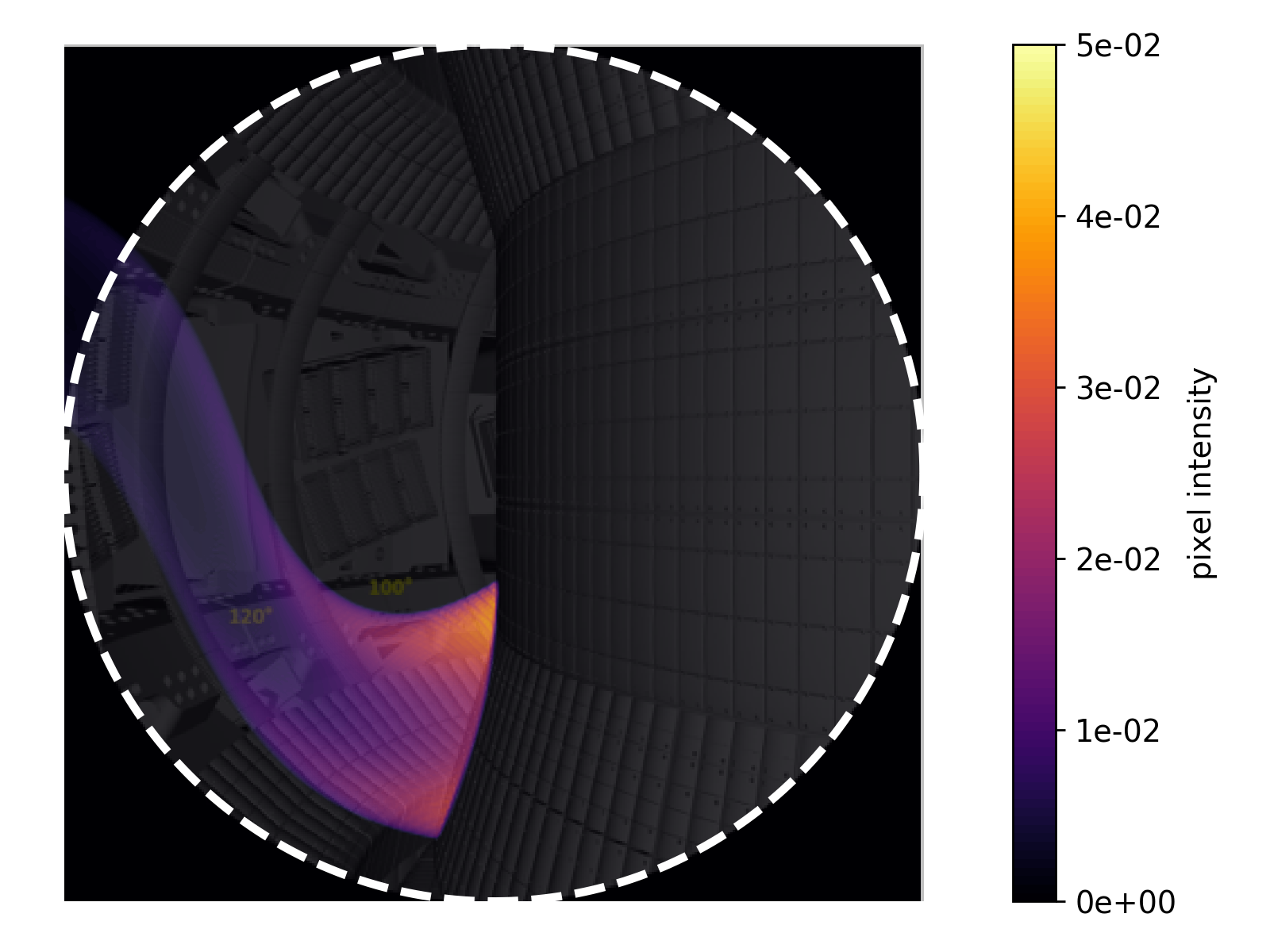}
            \caption{$\lambda = \SI{350-800}{nm}$}
            \label{fig:disruption_vis}
        \end{subfigure}
        \begin{subfigure}{\mywidth}
            \centering
            \includegraphics[height=\myheight]{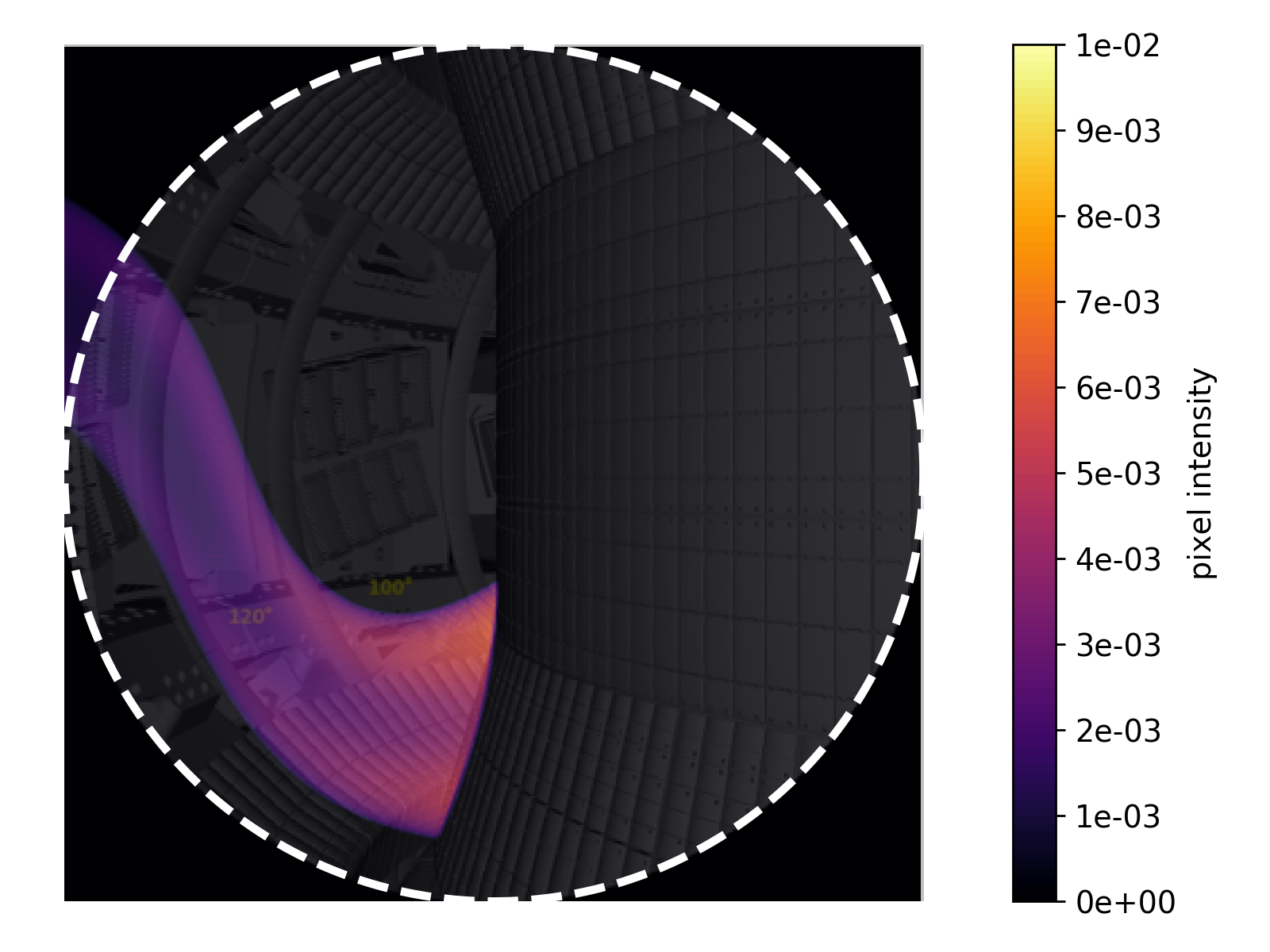}
            \caption{$\lambda = \SI{1.55-1.65}{\mu m}$}
            \label{fig:disruption_swir}
        \end{subfigure}
        \begin{subfigure}{\mywidth}
            \centering
            \includegraphics[height=\myheight]{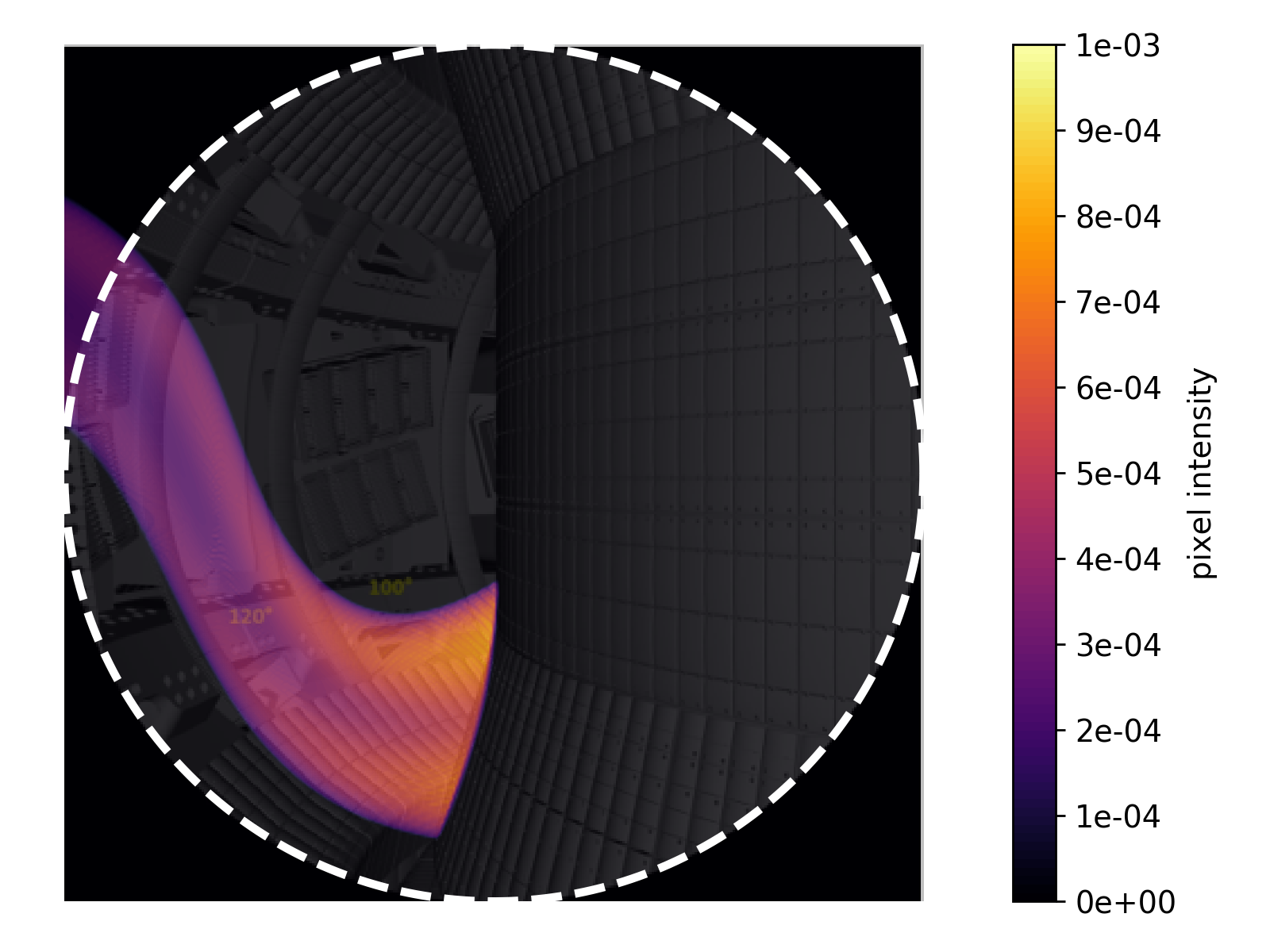}
            \caption{$\lambda = \SI{3.45-3.55}{\mu m}$}
            \label{fig:disruption_mwir}
        \end{subfigure}
        \caption{Synthetic camera images for REs populated with (a)~Gaussian radial distribution for the disruption equilibrium in \cref{fig:VDE-03cm}; throughput from \cref{fig:spectra_tp} is applied when integrating over spectral ranges (b)-(d). Colorscale is pixel intensity, normalized to the maximum integrated over $\lambda = \SI{0.35-4.1}{\mu m}$. FOV boundary is dashed. RE energy is ${\sim}\SI{30}{MeV}$, and pitch angle is uniform over $\thetap = \SI{0-0.1}{rad}$.} 
        \label{fig:disruption_gauss_spec}
    \end{figure}

    As expected from \cref{fig:spectra_tp}, the visible-range synthetic image (\cref{fig:disruption_vis}) is ${\sim}5$ times brighter than that for the short-wavelength IR range (\cref{fig:disruption_swir}), which is itself ${\sim}10$ times brighter than the mid-wavelength IR range (\cref{fig:disruption_mwir}). Of course, the true pixel intensity will depend on the true energy distribution function, but this is promising as SPARC plans to have high-speed visible cameras (${>}\SI{10\mathrm{k}}{fps}$) which should be capable of resolving disruption RE dynamics.
\section{Summary}\label{sec:summary}

    Matched co- and counter-$\Ip$ wide-angle views are planned for the SPARC main chamber (see \cref{fig:camera-views,tab:cam_params}), enabling forward-directed, visible and infrared (IR) synchrotron emission from runaway electrons (REs) to be detected by one view and background light to be subtracted using the other. As RE energies increase, their synchrotron spectrum shifts toward shorter wavelengths (see \cref{fig:spectra_tp}); thus, the expected light transmission of proposed in-vessel optics may limit detection to wavelengths $\lambda < \SI{4.1}{\mu m}$ and hence RE energies ${>}\SI{5}{MeV}$.

    The code SOFT is used to launch and follow REs during start-up, flat-top, and disruption phases of the plasma discharge (see \cref{fig:startup_flattop_eq,fig:VDE_eq}) and simulate their emitted and detected synchrotron radiation (see \cref{fig:flattop_all,fig:startup_bands_spec,fig:disr_bands,fig:disruption_gauss_spec}). Importantly, the proposed views can ``see'' REs from the plasma core to edge in all explored scenarios, except when the plasma drifts beyond $|Z_0| > \SI{0.4}{m}$.% (see \cref{fig:disr_bands}). 

    As anticipated, an IR camera is best suited to detect start-up REs, and frame rates $O(\SI{100}{fps})$ should be sufficient. However, hard x-ray monitors are needed to ``catch'' REs earlier at energies ${<}\SI{5}{MeV}$. On the other hand, a high-speed visible camera is a good choice for high-energy disruption REs. Future work will need to assess the absolute spectral radiance from more realistic RE distribution functions, as well as account for ex-vessel optics and true detector specifications.

\begin{acknowledgments}
    Thanks to C.~Clauser, D.~Battaglia, and M.~Hoppe. Work supported by Commonwealth Fusion Systems\rsiadd{: MIT RPP021 and RPP031}.
\end{acknowledgments}
%\input{contributions}

%\newpage
\bibliography{bib}

\end{document}